\documentstyle[sprocl]{article}
\catcode`\@=11
     
\def\ps@ppt{\def\@oddhead{\hfil FERMILAB--CONF--97/157--T}\def\@evenhead{\hfil 
FERMILAB--CONF--97/157--T}}     
\def\ps@titleppt{
 \def\@oddfoot{\hfill{{\footnotesize \textit{Lectures given at the Advanced School on 
 Electroweak Theory, Ma\'{o}, Menorca (June 1996).}}}\hfill}%

\def\@oddhead{\hfil FERMILAB--CONF--97/157--T}}     
\catcode`\@=12
     
\relax

\input boxedeps
\SetRokickiEPSFSpecial
\HideDisplacementBoxes

\def\pl#1#2#3{\frenchspacing{\it Phys. Lett. }{\bf #1}, #2 (19#3)}
        
\def\prl#1#2#3{\frenchspacing{\it Phys. Rev. Lett. }{\bf #1}, #2 (19#3)}

\def\rmp#1#2#3{\frenchspacing{\it Rev. Mod. Phys. }{\bf #1}, #2 (19#3)}

\def\prep#1#2#3{\frenchspacing{\it Phys. Rep. }{\bf #1}, #2 (19#3)}

\def\pr#1#2#3{\frenchspacing{\it Phys. Rev. D}{\bf #1}, #2 (19#3)}

\def\np#1#2#3{\frenchspacing{\it Nucl. Phys. }{\bf #1}, #2 (19#3)}

\def\zp#1#2#3{\frenchspacing{\it Z.~Phys. C}{\bf#1}, #2 (19#3)}

\def\arnps#1#2#3{\frenchspacing{\it Ann. Rev. Nucl. Part. Sci. }{\bf #1}, #2 (19#3)}

\def\ib#1#2#3{\frenchspacing{\it ibid. }{\bf #1}, #2 (19#3)}

\def\app#1#2#3{\frenchspacing{\it Acta Phys. Polon. B}{\bf #1}, #2 (19#3)}

\def\jmp#1#2#3{\frenchspacing{\it J. Math. Phys. }{\bf #1}, #2 (19#3)}

\def\npbps#1#2#3{{\em Nucl. Phys. B (Proc. Supp.)\/} {\bf #1} (19#3) #2}

\def\sciam#1#2#3#4{\frenchspacing{\it Sci. Am. }{\bf #1}, (\ifcase#3\or January\or February\or March\or April\or May\or
June\or July\or August\or September\or October\or November\or 
December\fi, 19#4), p.~#2}

\def\yadfiz#1#2#3#4#5#6{\frenchspacing{\it Yad. Fiz. }{\bf #1}, #2 (19#3) [English translation: 
         {\frenchspacing \it Sov. J. Nucl. Phys. }{\bf #4}, #5 (19#6)]}

\def\jetpl#1#2#3#4#5#6{\frenchspacing{\it ZhETF Pis'ma }{\bf #1}, #2 (19#3) [English translation: 
         {\frenchspacing \it JETP Lett. }{\bf #4}, #5 (19#6)]}
         
\def\mpl#1#2#3{{\it Mod. Phys. Lett. }{\bf #1}, #2 (19#3)}
         
\def\phystoday#1#2#3#4{\frenchspacing{\it Phys. Today }{\bf #1}, #2 (\ifcase#3\or January\or 
         February\or March\or April\or May\or June\or July\or August\or 
         September\or October\or November\or December\fi, 19#4)}
\newcommand{\hepph}[1]{(electronic archive: hep--ph/#1)}
\newcommand{\hepth}[1]{(electronic archive: hep--th/#1)}
\newcommand{\heplat}[1]{(electronic archive: hep--lat/#1)}
\newcommand{\hepex}[1]{(electronic archive: hep--ex/#1)}

\def\worldsci#1{(World Scientific, Singapore, 19#1)}

\relax

\def\slashii#1{\setbox0=\hbox{$#1$}             
   \dimen0=\wd0                                 
   \setbox1=\hbox{\sl/} \dimen1=\wd1            
   \ifdim\dimen0>\dimen1                        
      \rlap{\hbox to \dimen0{\hfil\sl/\hfil}}   
      #1                                        
   \else                                        
      \rlap{\hbox to \dimen1{\hfil$#1$\hfil}}   
      \hbox{\sl/}                               
   \fi}                                         %
\def\slashiii#1{\setbox0=\hbox{$#1$}#1\hskip-\wd0\hbox to\wd0{\hss\sl/\/\hss}}
\def\vev#1{\langle #1\rangle_0}
\def\avg#1{\langle #1\rangle}

\def\abs#1{\left| #1\right|}

\def\ltap{\mathop{\raisebox{-.4ex}{\rlap{$\sim$}} 
\raisebox{.4ex}{$<$}}}
\def\gtap{\mathop{\raisebox{-.4ex}{\rlap{$\sim$}} 
\raisebox{.4ex}{$>$}}}

\def\beq{\begin{equation}}
\def\eeq{\end{equation}}

\def\bea{\begin{eqnarray}}
\def\eea{\end{eqnarray}}

\def\bentarrow{\:\raisebox{1.3ex}{\rlap{$\vert$}}\!\rightarrow}

\def\twodk#1#2#3#4{
	\begin{equation}
	\begin{array}{l}
	#1\;#2 \\
	 \raisebox{1.3ex}{\rlap{$\vert$}}\raisebox{-0.5ex}{$\vert$}%
	\phantom{#2}\!\bentarrow #4\\
	\hspace{-2.2pt}\bentarrow #3 
	\end{array}
	\end{equation}
		}
\def\dkpp#1#2#3#4#5#6{
	\begin{equation}
	\begin{array}{r c l}
	#1 & \rightarrow & #2#3 \\
	 & & \phantom{\; #2}\bentarrow #4#5 \\
	 & & \phantom{\; #2 \bentarrow \;} \bentarrow #6
	\end{array}
	\end{equation}
		}
\newcommand{\etal}{{\em et al.}}
\newcommand{\ie}{{\em i.e.}}
\newcommand{\etc}{\emph{etc.}}

\newcommand{\vs}{\emph{vs.}\ }
\newcommand{\gevcc}{\hbox{ GeV}\!/\!c^2}
\newcommand{\gev}{\hbox{ GeV}}
\newcommand{\ev}{\hbox{ eV}}
\newcommand{\mev}{\hbox{ MeV}}
\newcommand{\mevcc}{\hbox{ MeV}\!/\!c^2}
\newcommand{\tev}{\hbox{ TeV}}
\newcommand{\tevcc}{\hbox{ TeV}\!/\!c^2}

\newcommand{\pb}{\hbox{ pb}}
\newcommand{\cm}{\hbox{ cm}}
\newcommand{\km}{\hbox{ km}}

\newcommand{\eqn}[1]{(\ref{#1})}
\newcommand{\cfrac}[2]{\textstyle \frac{#1}{#2}}
\def\onetev{1-TeV scale}
\newcommand{\ewsb}{electroweak symmetry breaking}

\newcommand{\ys}{\hbox{ ys}}
\newcommand{\ps}{\hbox{ ps}}

\newcommand{\any}{\hbox{ anything}}
\def\ws{$SU(2)_L\otimes U(1)_Y$}
\def\sm{$SU(3)_c\otimes SU(2)_L\otimes U(1)_Y$}
\def\D{{\cal D}}
\def\L{{\cal L}}
\def\rr#1{[{\frenchspacing #1\nonfrenchspacing}]}

\def\fm{\hbox{ fm}}
\def\ubar{\overline{u}}
\def\dbar{\overline{d}}
\def\sbar{\overline{s}}
\def\shat{{\hat{s}}}
\def\that{{\hat{t}}}
\def\uhat{{\hat{u}}}

\def\qbar{\overline{q}}

\newcommand{\diamant}{\BoxedEPSF{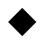  scaled 750}}

\def\beq{\begin{equation}}
\def\eeq{\end{equation}}
\def\bea{\begin{eqnarray}}
\def\eea{\end{eqnarray}}

\begin{document}
\title{HADRON COLLIDERS, THE TOP QUARK,\\ AND THE HIGGS SECTOR}

\author{CHRIS QUIGG}

\address{Fermi National Accelerator Laboratory \\ P. O. Box 500, 
Batavia, Illinois 60510 USA \\ and \\ Department of Physics, Princeton 
University \\ Princeton, New Jersey 08540 USA}


\maketitle\abstracts{
I survey the characteristics of hadron colliders as tools to 
investigate top-quark physics and to explore the \onetev\ of 
electroweak symmetry breaking.}

\section{Hadron Colliders for the 1-TeV Scale}
Hadron colliders respond to our need to study a rich diversity of elementary
processes at high energies.  For orientation, let us suppose that we 
wish to study quark-quark collisions at a c.m.\ energy of $1\tev$.  
If we say that three quarks share one-half of a proton's momentum, 
\ie, $\avg{x} = \cfrac{1}{6}$ for a quark, then we require 
proton-proton collisions at $\sqrt{s} \approx 6\tev$. 

To achieve this c.m.\ energy in a fixed-target machine would require 
a proton beam momentum $p \approx 2 \times 10^{4}\tev = 2 \times 
10^{16}\ev$, which approaches the momentum ($10^{19}\hbox{ to 
}10^{20}\ev$) of the highest-energy cosmic rays.  The radius of a ring 
that magnetically confines a beam of momentum $p$ is
\begin{equation}
	r  =  \frac{10}{3} \cdot \left( \frac{p}{1\tev}\right) / 
	\left( \frac{B}{1\hbox{ tesla}}\right)\km .
	\label{radius}
\end{equation}
To confine a beam with $p = 2\times 10^{4}\tev$ in conventional copper 
magnets with a field of $B = 2\hbox{ teslas}$ would require a radius of
\begin{equation}
	r \approx \cfrac{1}{3} \times 10^{5}\km.
	\label{rad6}
\end{equation}
The distance from Earth to the Moon is about $4 \times 10^{5}\km$, so 
we see that the radius \eqn{rad6} is about $\cfrac{1}{12}$ the size 
of the Moon's orbit.  A conventional fixed-target machine to explore 
the \onetev\ is a very large accelerator indeed!  Superconducting 
magnets help---but not enough: a 10-tesla field reduces the size of 
the accelerator to mere Earth size ($R_{\oplus} = 6.4 \times 
10^{3}\km$).

The solution---one of the great technological achievements of 
accelerator physics---is to collide one high-energy beam with 
another.  Then to reach $\sqrt{s} = 6\tev$, we need two 3-TeV proton 
synchrotrons, with radius
\begin{equation}
	r_{3} = \frac{10\hbox{ teslas}}{B}\km.
	\label{rho3}
\end{equation}
For a 5-T magnetic field, well within current practice, the ring need 
only have a radius of $r_{3} \approx 2\km$.  Within some factor 
determined by the physics studies to be pursued in detail, this 
estimate defines the natural scale for a hadron collider to explore 
the \onetev. Because we have considered only quark-quark collisions, 
and have treated the proton very simply, this is a very rough 
estimate.  Quark-antiquark, or glue-glue, or $WW$ collisions at 
$1\tev$ will require a higher energy $pp$ machine.  An estimate based 
on a circular ring is also an underestimate because accelerators need 
space for focusing magnets and for experiments---our reason for 
building the accelerators in the first place!
\subsection{Hadron Colliders through the Ages}
The first hadron collider was the CERN Intersecting Storage Rings, 
which came into operation around 1970.  The ISR was a $pp$ collider 
that eventually reached $\sqrt{s} = 63\gev$.  It was composed of two 
rings of conventional magnets.

About a decade later, the SPS Collider came into operation at CERN.  
Counter-rotating beams of protons and antiprotons were confined in a 
single conventional-magnet synchrotron (the CERN SPS) to provide 
collisions at $\sqrt{s}=630\gev$.  The S$\bar{p}p$S Collider was home 
to the famous UA1 and UA2 detectors and a number of specialized 
experiments.

The Fermilab Tevatron, the first superconducting synchrotron, was 
commissioned in 1983.  Like the S$\bar{p}p$S, the Tevatron is 
a proton-antiproton collider, with 900-GeV beams confined by 4-T 
magnets.  The general-purpose CDF and D\O\ detectors and several 
specialized experiments have been installed in the Tevatron tunnel.  
In 1999, the energy of the Tevatron will be raised to $1\tev$ per 
beam by cooling the magnets below 4 kelvins.

The Superconducting Super Collider (SSC) was to have been a $\sqrt{s}=40\tev$ 
proton-proton collider.  With 6.6-T magnets, its circumference was 
$87\km$.  Alas, the United States abandoned construction in 1993.

The next machine on the horizon is the Large Hadron Collider (LHC) at 
CERN, a proton-proton collider under construction in the 27-km LEP 
tunnel.  The LHC will be built of 9-T superconducting magnets 
operating at about $1.8\hbox{ K}$, the temperature of superfluid 
helium.  The LHC is our great hope for exploring the \onetev\ and 
unraveling the puzzle of electroweak symmetry breaking.  Experiments 
may begin as early as 2005.

I should mention as well that the Relativistic Heavy Ion Collider (RHIC) at 
Brookhaven will have for part of each year (polarized) proton-proton collisions 
at $\sqrt{s}=400\gev$.  RHIC is a pair of superconducting rings with 
modest magnetic field.

\subsection{Key Advances in Accelerator Technology}
\begin{figure}[tb]
	\centerline{\BoxedEPSF{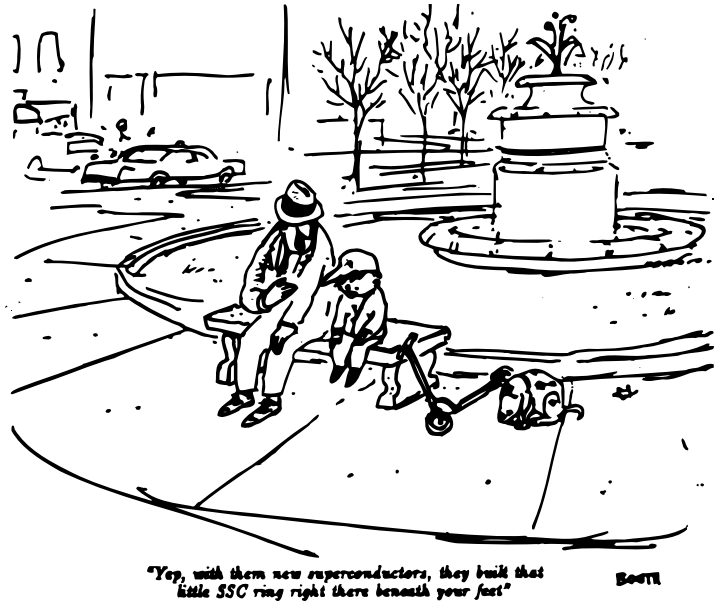  scaled 550}}
	\caption{Illustration for the poster advertising a talk on 
	high-$T_{c}$ superconductors at the SSC Central Design Group.}
	\protect\label{fig:hiTc}
\end{figure}

Our ability to contemplate experiments at very high energies is owed 
to many important advances in accelerator technology.  I would list 
six as being of defining importance:
\begin{itemize}
	\item  The idea of colliding beams.

	\item  Alternating-gradient (``strong'') focusing, invented by 
	Christofilos, Cour\-ant, Livingston, and Snyder.

	\item  Superconducting accelerator magnets.  We owe to materials 
	scientists the discovery of practical ``type-II'' superconductors, 
	including the NbTi used in all superconducting machines to date, and 
	the brittle Nb$_{3}$Sn, which may find use in special applications.  
	The superconducting cable used in accelerator magnets has roots in 
	pioneering work carried out at the Rutherford Laboratory, and 
	essential early steps in the development of robust magnet structures 
	were taken at Fermilab.

	\item  The evolution of vacuum technology.  Accelerator beams stored 
	for approximately 20 hours must travel approximately $2\times 
	10^{10}\km$, about 150 times the distance from Earth to Sun, without 
	encountering a stray air molecule.

	\item  The development of large-scale cryogenic technology, to 
	maintain many kilometers of magnets at a few kelvins.

	\item  We owe the S$\bar{p}p$S and Tevatron colliders to the 
	development of intense antiproton sources, building on the work of 
	Budker at Novosibirsk and van der Meer at CERN.
\end{itemize}
We would of course be nowhere without dreams and fantasies.  I show 
in Figure \ref{fig:hiTc} an artifact from my days in the SSC Central 
Design Group in Berkeley.

\subsection{Competing Technologies}
There are no competing technologies, broadly speaking, if we wish to 
study collisions of quarks and gluons.  There are, of course, many 
design choices to make, within the general framework of a hadron 
collider: $pp$ \vs $\bar{p}p$, low-field magnets \vs high-field 
magnets, \etc  

For the study of lepton-lepton scattering, LEP is the shining example 
of a (reasonably) high-energy $e^{+}e^{-}$ collider.  It is widely 
agreed that the rise of synchrotron radiation causes circular electron 
machines to become impractical for energies above a few hundred GeV.  
Linear colliders, discussed at this school by Ram\'{o}n 
Miquel,\cite{miquel} are therefore under development for $\sqrt{s}$ 
from a few hundred GeV to about $1.5\tev$.  The central question for 
linear colliders is, what performance at what cost?  I think it possible that 
linear-collider technology may only be interesting for about one 
decade in energy; the growth path beyond $1\hbox{ to }2\tev$ is not 
clear.  But it is a very interesting decade, the one on which we 
expect to learn the secrets of \ewsb, which is why there is such 
intense interest in the technology.

Over the past few years, the possibility of a muon collider has 
received increased attention.\cite{mumu}  The muon's brief lifetime (2.2 $\mu$s 
at rest) means that one imperative is to move fast!  If a muon 
collider is practical, it may be attractive for lepton-lepton 
collisions at far higher energies than we contemplate for 
electron-positron machines, since the greater mass of the muon 
($m_{\mu}\gg m_{e}$) means that synchrotron radiation is a minor 
concern.  The muon collider is the least developed technology of the 
hadron, electron, and muon alternatives, but it is not obviously 
impossible.

\section{What Landmarks Do We Expect?}{\protect{\label{landm}}}
We have already remarked in the Introduction on the importance of
the \onetev.\cite{ehlq} In this section, we wish to review for the first
time some of the arguments that lead to an identification of the
\onetev\ as a key landmark. As we shall see again and again in different
ways, our understanding of the spontaneous breaking of the electroweak
gauge symmetry is incomplete. A more complete understanding can be
obtained only with the aid of a thorough knowledge of what takes
place on the \onetev.

Let us review the essential elements of the \ws\ electroweak 
theory.\cite{GT}
To save writing, we shall
speak of the model as it applies to a single generation of leptons.
In this form, the model is neither complete nor consistent: anomaly
cancellation requires that a doublet of color-triplet quarks accompany
each doublet of color-singlet leptons. However, the needed
generalizations are simple enough to make that we need not write
them out.

We begin by specifying the fermions: a left-handed weak isospin doublet
\beq
{{\sf L}} = \left(\begin{array}{c} \nu_e \\ e \end{array}\right)_L
\eeq
with weak hypercharge $Y_L=-1$, and a right-handed weak isospin singlet
\beq
      {{\sf R}}\equiv e_R
\eeq
with weak hypercharge $Y_R=-2$.

The electroweak gauge group, \ws, implies two sets of gauge fields:
a weak isovector $\vec{b}_\mu$, with coupling constant $g$, and a
weak isoscalar
${{\mathcal A}}_\mu$, with coupling constant $g^\prime$. Corresponding
to these gauge fields are the field-strength tensors $\vec{F}_{\mu\nu}$
for the weak-isospin symmetry and $f_{\mu\nu}$ for the weak-hypercharge
symmetry.

We may summarize the interactions by the Lagrangian
\beq
\L = \L_{\rm gauge} + \L_{\rm leptons} \ ,                           
\eeq             
with
\beq
\L_{\rm gauge}=-\frac{1}{4}F_{\mu\nu}^\ell F^{\ell\mu\nu}
-\frac{1}{4}f_{\mu\nu}f^{\mu\nu},
\eeq
and
\begin{eqnarray}     
\L_{\rm leptons} & = & \overline{{\sf R}}\:i\gamma^\mu\!\left(\partial_\mu
+i\frac{g^\prime}{2}{\cal A}_\mu Y\right)\!{\sf R} \\ & + & \overline{{\sf
L}}\:i\gamma^\mu\!\left(\partial_\mu 
+i\frac{g^\prime}{2}{\cal
A}_\mu Y+i\frac{g}{2}\vec{\tau}\cdot\vec{b}_\mu\right)\!{\sf L}. \nonumber
\end{eqnarray}

To hide the electroweak symmetry, we introduce a complex doublet
of scalar fields
\beq
\phi\equiv \left(\begin{array}{c} \phi^+ \\ \phi^0 \end{array}\right)
\eeq
with weak hypercharge $Y_\phi=+1$. Add to the Lagrangian new terms
for the interaction and propagation of the scalars,
\beq
      \L_{\rm scalar} = (\D^\mu\phi)^\dagger(\D_\mu\phi) - V(\phi^\dagger \phi),
\eeq
where the gauge-covariant derivative is
\beq
      \D_\mu=\partial_\mu 
+i\frac{g^\prime}{2}{\cal A}_\mu
Y+i\frac{g}{2}\vec{\tau}\cdot\vec{b}_\mu \; ,
\eeq
and the potential interaction has the form
\beq
      V(\phi^\dagger \phi) = \mu^2(\phi^\dagger \phi) +
\abs{\lambda}(\phi^\dagger \phi)^2 .
\label{SSBpot}
\eeq
We are also free to add a Yukawa interaction between the scalar fields
and the leptons:
\beq
      \L_{\rm Yukawa} = -G_e\left[\overline{{\sf R}}(\phi^\dagger{\sf
L}) + (\overline{{\sf L}}\phi){\sf R}\right].
\eeq

The electroweak symmetry is spontaneously broken if the parameter
$\mu^2<0$. The minimum energy, or vacuum state, may then be chosen
to correspond to the vacuum expectation value
\beq
\vev{\phi} = \left(\begin{array}{c} 0 \\ v/\sqrt{2} \end{array}
\right),
\eeq
where
\begin{eqnarray}
      v = \sqrt{-\mu^2/\abs{\lambda}} & = &
\left(G_F\sqrt{2}\right)^{-\frac{1}{2}} \\
 & \approx & 246~\gev \nonumber
\end{eqnarray}
is fixed by the low-energy phenomenology of charged current
interactions.

The spontaneous symmetry breaking has several important consequences:
\begin{itemize}
\item Electromagnetism is mediated by a massless photon, coupled to the electric
charge;
\item The mediator of the charged-current weak interaction acquires
a mass characterized by $M_W^2=\pi\alpha/G_F\sqrt{2}\sin^2{\theta_W}$,
where $\theta_W$ is the weak mixing angle;
\item The mediator of the neutral-current weak interaction acquires
a mass characterized by $M_Z^2=M_W^2/\cos^2{\theta_W}$;
\item A massive neutral scalar particle, the Higgs boson, appears,
but its mass is not predicted;
\item Fermions (the electron in our abbreviated treatment) can acquire
mass.
\end{itemize}

It is well known that the Standard Model does not give a precise 
prediction for the mass of the Higgs boson. We can, however, use arguments 
of self-consistency to place plausible lower and upper bounds on the mass of 
the Higgs particle in the minimal model. A lower bound is obtained by 
computing~\cite{SSI18} the first quantum corrections to the classical potential
\eqn{SSBpot}. Requiring that $\vev{\phi}\neq 0$ be an absolute minimum of the one-loop 
potential yields the condition
\begin{eqnarray}
	M_H^2 & > & 3G_F\sqrt{2}(2M_W^4+M_Z^4)/16\pi^2 \\
	 & \gtap & 7~\gev/c^2 \nonumber \; .
\end{eqnarray}

Unitarity arguments~\cite{lqt} lead to a conditional upper bound on the Higgs 
boson mass. It is straightforward to compute the 
amplitudes ${\cal M}$ for gauge boson scattering at high energies, and to make
a partial-wave decomposition, according to
\beq
      {\cal M}(s,t)=16\pi\sum_J(2J+1)a_J(s)P_J(\cos{\theta}) \; .
\eeq
 Most channels ``decouple,'' in the sense 
that partial-wave amplitudes are small at all energies (except very
near the particle poles, or at exponentially large energies), for
any value of the Higgs boson mass $M_H$. Four channels are interesting:
\beq
\begin{array}{cccc}
W_L^+W_L^- & Z_L^0Z_L^0/\sqrt{2} & HH/\sqrt{2} & HZ_L^0 \; ,
\end{array}
\eeq
where the subscript $L$ denotes the longitudinal polarization
states, and the factors of $\sqrt{2}$ account for identical particle
statistics. For these, the $s$-wave amplitudes are all asymptotically
constant (\ie, well-behaved) and  
proportional to $G_FM_H^2.$ In the high-energy limit, 
\beq
\lim_{s\gg M_H^2}(a_0)\to\frac{-G_F M_H^2}{4\pi\sqrt{2}}\cdot \left[
\begin{array}{cccc} 1 & 1/\sqrt{8} & 1/\sqrt{8} & 0 \\
      1/\sqrt{8} & 3/4 & 1/4 & 0 \\
      1/\sqrt{8} & 1/4 & 3/4 & 0 \\
      0 & 0 & 0 & 1/2 \end{array} \right] \; .
\eeq 
Requiring that the largest eigenvalue respect the 
partial-wave unitarity condition $\abs{a_0}\le 1$ yields
\beq
	M_H \le \left(\frac{8\pi\sqrt{2}}{3G_F}\right)^{1/2} =1\tevcc
\eeq
as a condition for perturbative unitarity.

If the bound is respected, weak interactions remain weak at all
energies, and perturbation theory is everywhere reliable. If the
bound is violated, perturbation theory breaks down, and weak
interactions among $W^\pm$, $Z$, and $H$ become strong on the \onetev.
This means that the features of strong interactions at \gev\ energies
will come to characterize electroweak gauge boson interactions at
\tev\ energies. We interpret this to mean that new phenomena are to
be found in the electroweak interactions at energies not much larger
than 1~TeV.

It is worthwhile to note in passing that 
the threshold behavior of the partial-wave amplitudes for gauge-boson 
scattering follows generally from chiral symmetry.\cite{LT8}  The partial-wave 
amplitudes $a_{IJ}$ of definite isospin $I$ and angular momentum $J$ are 
given by
\begin{eqnarray}
	a_{00} \approx & G_Fs/8\pi\sqrt{2} & \hbox{attractive,}
	\nonumber \\
	a_{11}  \approx & G_Fs/48\pi\sqrt{2} & \hbox{attractive,} \\
	a_{20} \approx & -G_Fs/16\pi\sqrt{2} & \hbox{repulsive.}
	\nonumber
\end{eqnarray} 

\begin{quotation}
	{\small \textbf{Problem: }Because the most serious high-energy divergences of 
	a spontaneously broken gauge theory are associated with the longitudinal 
	degrees of freedom of the gauge bosons, which arise from auxiliary 
	scalars, it is instructive to study the Higgs sector in isolation.  
	Consider, therefore, the Lagrangian for the Higgs sector of the 
	\sm\ electroweak theory before the gauge couplings are turned on,
	\begin{displaymath}
		\L_{\mathrm{scalar}} = (\partial^\mu \phi)^\dagger (\partial_\mu \phi) 
		-\mu^2 (\phi^\dagger \phi) - \abs{\lambda}(\phi^\dagger \phi)^2.
	\end{displaymath} \\
	(a) Choosing $\mu^2 < 0$, investigate the effect of spontaneous symmetry 
	breaking.  Show that the theory describes three massless scalars 
	$(w^+, w^-, z^0)$ and one massive neutral scalar $(h)$, which interact 
	according to
	 \begin{eqnarray*}
	 	\L_{\mathrm{int}} & = & -\abs{\lambda} v h (2w^+ w^- + z^2 + h^2)  \\
	 	 &  & -(\abs{\lambda}/4)(2w^+ w^- + z^2 + h^2)^2 ,
	 \end{eqnarray*} where $v^2 = -\mu^2/\abs{\lambda}$.  In the language of 
	 the full electroweak theory, $1/v^2 = G_F\sqrt{2}$ and 
	 $\lambda = G_FM_H^2/\sqrt{2}$. 
	 \\
	 (b) Deduce the Feynman rules for interactions and compute the 
	 lowest-order (tree diagram) amplitude for the reaction $hz \rightarrow 
	 hz$. \\
	 (c) Compute the $J=0$ partial-wave amplitude in the high-energy limit 
	 and show that it respects partial-wave unitarity only if 
	 $M_H^2 < 8\pi\sqrt{2}/G_F$. \rr{Reference: B. W. Lee, C. Quigg, and H. 
	 B. Thacker, \pr{16}{1519}{77}.} }
\end{quotation}

\section{What is a proton?}
For the construction of large accelerators, we are limited to beams
of charged, stable particles, which means to electrons and protons.
With current methods, it is feasible to produce intense proton beams of
tens of \tev, but electron beams of only about a tenth of a \tev.
So far as we know, the electron is an elementary point
particle ($r_e \ltap 10^{-16}\cm$), but the proton is a composite
system. Our ability to exploit the energy advantage of proton beams
therefore depends on our knowledge of what a proton is, and how it
behaves in high energy collisions. The purpose of this section is
to summarize the state of our knowledge.

The static properties of a proton are well characterized by a
description of a proton as a three-quark ($uud$) bound state, with
a radius $r_p\approx 1\fm$. This picture accounts for the essential
features of magnetic moments, axial charges, electromagnetic form
factors, and such.\cite{SSI12}

What we might call the quasistatic properties of a proton are 
attributes measured in hard-scattering processes, but determined by 
low-energy (nonperturbative) dynamics.  I have in mind the 
flavor-asymmetry of the light-quark sea, $u_{s}(x) \neq d_{s}(x)$, 
which can be understood from the chiral dynamics of constituent quarks 
and Goldstone bosons,\cite{ehq} and the spin structure of the proton.

In collision, especially for the purpose of hard scattering,
a proton is a broad-band, unselected beam of quarks, antiquarks,
and gluons,
and possibly other constituents as well. The composition of this
mixed beam depends on how you inspect it: the more virtual the probe,
the more sensitive it will be to short time-scale fluctuations.

It is fruitful to analyze the proton in the framework of the {\em
parton model} with QCD refinements. The fundamental quantity in this
picture is $f_i^{(a)}(x_a,Q^2)$, the number density of partons of
species $i$ with momentum fraction $x_a$ of hadron $a$ seen by a
probe with resolving power characterized by $Q^2$.

Up to now, the best information on parton distributions (or hadron
structure functions) comes from measurements of deeply inelastic
lepton scattering, the reactions
\begin{eqnarray}
eN & \to & e+\any, \\
\mu N & \to & \mu+\any,    \\
\nu_\mu N & \to & \mu+\any, \label{eq:cc}
\end{eqnarray}
and
\beq
\nu_\mu N \to \nu_\mu+\any. \label{eq:nc}
\eeq
For the scattering of charged leptons at present energies, the probe
is a virtual photon (with usually negligible corrections for the
exchange of a virtual $Z^0$). In the charged-current reaction 
\eqn{eq:cc},
the nucleon is probed by $W^\pm$; in the neutral-current reaction
\eqn{eq:nc}, the probe is the $Z^0$.

\begin{figure}[tb]
	\centerline{\BoxedEPSF{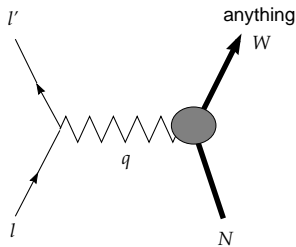  scaled 400}}
	\caption{Kinematics of deeply inelastic scattering.}
	\protect\label{fig:DIS}
\end{figure}

The kinematic notation for deeply inelastic scattering is indicated
in Figure \ref{fig:DIS}.
From the four-momenta indicated there we may form the useful invariants
\begin{eqnarray}
s & = & (\ell+ P)^2, \\
Q^2 & \equiv & -q^2 = -(\ell-\ell^\prime)^2, \\
\nu & = & q\cdot P/M,
\end{eqnarray}
where $M$ is the target mass, and
\beq
W^2=2M\nu+M^2-Q^2
\eeq
is the square of the invariant mass of the produced hadronic system
``\any.'' It is convenient to work in terms of Bjorken's dimensionless
variables
\beq
x=Q^2/2M\nu,
\eeq
the momentum fraction of the struck parton, and
\beq
y=\nu/E_{\rm lab},
\eeq
the fractional energy loss of the leptons in the laboratory frame.

For electromagnetic scattering, we may write the differential cross
section as
\beq
\frac{d\sigma}{dxdy}=\frac{4\pi\alpha^2s}{Q^4}\left[F_2(x)(1-y)+F_1(x)xy^2\right].
\eeq
In the parton model, $2xF_1(x)=F_2(x)$, and the structure function $F_2$ of the proton may
be written as
\beq
F_2^{ep}(x)/x =
\cfrac{4}{9}\left(\vphantom{d}u(x)+\ubar(x)\right)+\cfrac{1}{9}\left(d(x)+\dbar(x)\right)
+\cfrac{1}{9}\left(\vphantom{d}s(x)+\sbar(x)\right)+\ldots
\eeq
The structure function of the neutron is obtained by an isospin
rotation, which is to say, by the replacement $u\leftrightarrow d$.
The parton distributions satisfy the momentum sum rule,
\beq
\sum_{i={\rm parton \atop species}}\int_0^1dx x f_i(x) = 1.
\label{eq:sr}
\eeq
An important early result was the recognition that charged partons
do not carry all the momentum of the nucleon. We may see this by
approximating
\beq
F_2^{ep}(x)+F_2^{en}(x) =
\cfrac{5}{9}x\left(u(x)+\ubar(x)+d(x)+\dbar(x)\right).
\eeq
A measurement of $F_2$ then leads to an estimate of the momentum
carried by charged partons through the connection
\begin{eqnarray}
\cfrac{9}{5}\int_0^1dx\left(F_2^{ep}(x)+F_2^{en}(x)\right) & = &
\sum_{i={\rm {quarks +} \atop antiquarks}}\int_0^1 dx x f_i(x) \\
 & = & 0.45~{\rm experimental}.\nonumber
\end{eqnarray}
Unless most of the momentum of the nucleon is carried by strange
(and heavier) quarks, this implies that about half the momentum of
a proton is carried by neutrals. 

Charged-current scattering of neutrinos from nucleons has also been
studied extensively. We define an ``isoscalar nucleon''
$N\equiv\frac{1}{2}(p+n)$. The differential cross sections for
scattering of neutrinos and antineutrinos are then
\begin{eqnarray}
\frac{d\sigma(\nu N\to \mu^-+X)}{dxdy} & = &
\frac{G_F^2ME}{\pi}\left[\left(u(x)+d(x)\right) + 
\left(\ubar(x)+\dbar(x)\right)(1-y)^2\right], \nonumber \\
 & & \\
\frac{d\sigma(\overline{\nu} N\to \mu^++X)}{dxdy} & = & \frac{G_F^2ME}{\pi}
\left[\left(u(x)+d(x)\right)(1-y)^2 +
\left(\ubar(x)+\dbar(x)\right)\right].\nonumber \\
 & & 
\end{eqnarray}
The difference $\sigma(\nu N)-\sigma(\overline{\nu}N)$ allows a
determination of the excess of quarks over antiquarks, \ie, the
distribution of ``valence'' quarks that determine the nucleon quantum numbers:
\beq
\Delta(x) \equiv u(x)-\ubar(x)+d(x)-\dbar(x)\equiv u_{\mathit{valence}}+d_{\mathit{valence}}.
\eeq

\begin{figure}[tb]
	\centerline{\BoxedEPSF{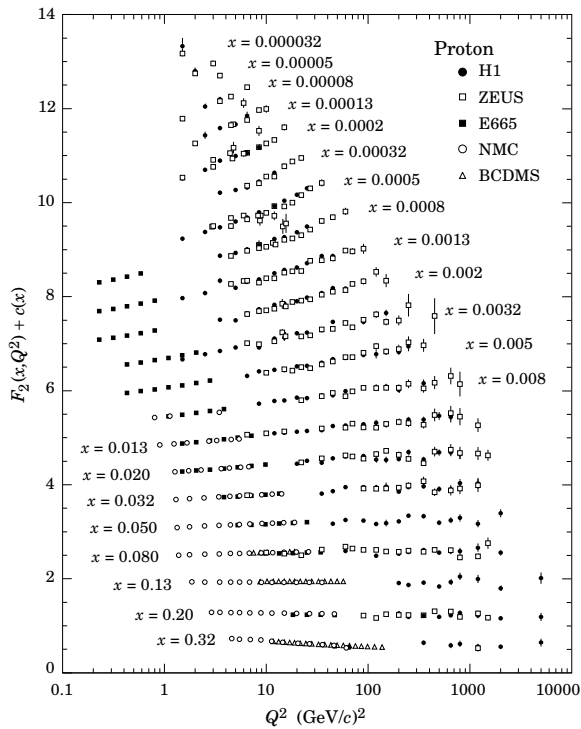  scaled 750}}
	\caption{The proton structure function $F_{2}^{p}$ measured in 
	electron (H1, ZEUS) and muon (BCDMS, E665, NMC) scattering [from the 1996
	 \textit{Review of Particle Physics} 
	{\protect\cite{pdg96}}].}
	\protect\label{fig:f2}
\end{figure}

Viewed at very long wavelengths, the proton appears structureless,
but as $Q^2$ increases and the resolution becomes finer, the proton
is revealed as a composite object characterized, for example, by
rapidly falling elastic form factors that decrease as $1/Q^4$. According
to the parton model, which ignores interactions among the constituents
of the proton, the picture for deeply inelastic scattering is then
exceedingly simple. Once $Q^2$ has become
large enough for the quark constituents to be resolved, no finer
structure is seen. The quarks are structureless, have no size, and
thus introduce no length scale. When $Q^2$ exceeds a few $\gev^2$,
all fixed mass scales become irrelevant and the structure functions and
parton distributions 
do not depend upon $Q^2$. That this is approximately so in Nature
may be seen from the measurements of $F_2^{p}(x,Q^{2})$ shown in 
Figure \ref{fig:f2}.

In an interacting field theory, however, a more complex picture of
hadron structure emerges. As $Q^2$ increases beyond the magnitude
required to resolve quarks, the quarks themselves are found to have
an apparent structure, which arises from the interactions mediated
by the gluon fields. The parton distributions
{\it evolve} with $Q^2$ as a result of quantum fluctuations. The 
virtual dissociation of a quark into a quark and gluon
degrades the valence quark distribution.
The virtual dissociation of a gluon into a $q\overline{q}$ pair
enhances the population of quarks and antiquarks.

\begin{figure}[tb]
	\centerline{\BoxedEPSF{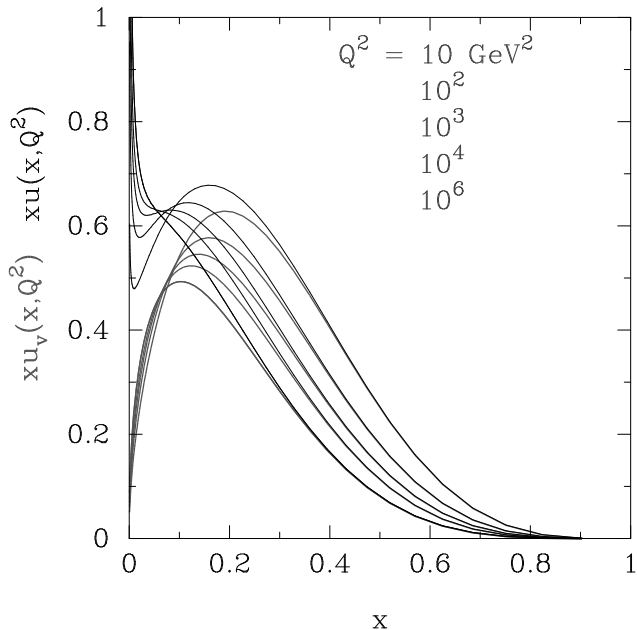  scaled 500}}
	\caption{The up-quark distribution $xu(x,Q^{2})$ in the proton for 
	several values of $Q^{2}$, according to the CTEQ4 parton distribution 
	functions.}
	\protect\label{fig:xu}
\end{figure}

It is therefore plausible to expect, in any interacting field theory,
that as $Q^2$ increases the structure function will fall at large
values of $x$ and rise at small values of $x$.
In most field theories, there is a power-law dependence on
$Q^2$, but in asymptotically free gauge theories such as QCD, the dependence on
$Q^2$ is only logarithmic. The $Q^{2}$-evolution of the up-quark 
distribution $xu(x,Q^{2})$ in the CTEQ4 parton 
distributions~\cite{cteq4} is shown in Figure \ref{fig:xu}.

The flavor ``nonsinglet''
structure function
\beq
x{\cal{F}}_3^{\nu N} = x\left[u(x)-\ubar(x)+d(x)-\dbar(x)\right],
\eeq
measures the valence quark distribution.  It is of special
interest because it receives no contribution from the dissociation
of gluons into quark-antiquark pairs; it is simply degraded, with
increasing $Q^2,$ by gluon radiation from the valence quarks. It
therefore offers, in principle, a means for studying the evolution
of the quark distributions uncomplicated by the need to know anything
about the gluon distribution. 

Once parton distributions have been measured in detail at some value
of $Q^2=Q_0^2,$ and the running coupling constant $\alpha_s(Q^2)$
of the strong interactions has been determined, QCD permits us to
compute the parton distributions at higher values of $Q^2$. A convenient
formalism is provided by the Altarelli-Parisi equations,\cite{SSI15}
integro-differential equations for the parton distributions.\cite{SSI16}
It is worth recalling a few of the essentials here.

It is conventional to parameterize the strong coupling constant as
\beq
1/\alpha_s(Q^2) = \frac{33-2n_f}{12\pi}\ln(Q^2/\Lambda^2),
\eeq
where $n_f$ is the number of ``active'' quark flavors, and to determine $\Lambda$ from the evolution of structure functions.
For example, if we define the second moment 
\beq
\Delta_2(Q^2) = \int_0^1 dx x \left[u_v(x)+d_v(x)\right]
\eeq
of the valence quark distribution, then the Altarelli-Parisi equations
give
\beq
\frac{\Delta_2(Q^2)}{\Delta_2(Q_0^2)} =
\left[\frac{\alpha_s(Q_0^2)}{\alpha_s(Q^2)} \right]^{6A_2/(33-2n_f)},
\eeq
where $A_2\simeq1.78.$ Knowing $\int_0^1 dx x
{\cal{F}}_3(x,Q^2)$ at different values of $Q^2$ thus allows, in
principle, a direct determination of the QCD scale parameter $\Lambda$.
In practice, the limited 
statistics of neutrino experiments, the small available range in
$\ln(Q^2)$, and other factors limit the precision of such
determinations.

If the evolution of $F_2^{\mu N}$ or ${\cal{F}}_2^{\nu N}$, either
of which is measured with higher statistics than $x{\cal{F}}_3^{\nu N}$,
is to be used for a determination of $\Lambda$, we are faced with
the problem that the gluon distribution is not measured directly
in lepton scattering. Its character must be inferred from the behavior 
of the antiquark distribution, and so is coupled with
the value of $\Lambda$.
\begin{figure}[tb]
	\centerline{\BoxedEPSF{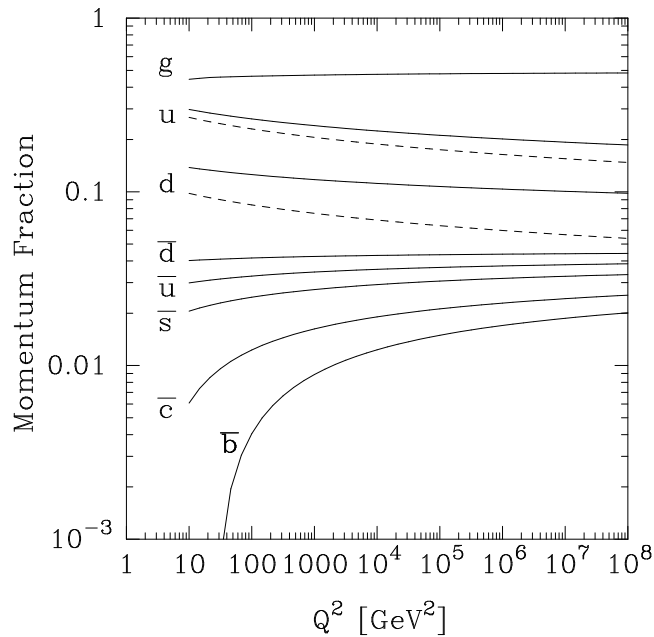  scaled 500}}
	\caption{$Q^{2}$-evolution of the momentum fractions carried by various 
	parton species.}
	\protect\label{fig:momfrac}
\end{figure}

As a final response to our question, ``What is a proton?''
let us look at the flavor content of the proton, as measured by the
momentum fraction
\beq
      \int_0^1 dx xf_i(x,Q^2)
\eeq
carried by each parton species. This is shown in Figure \ref{fig:momfrac} 
for the CTEQ4 parton distributions.\cite{cteq4}
As $Q^2$ increases, momentum is shared more and
more equally among the quark and antiquark flavors, reflecting the
trend toward the asymptotic values
\begin{eqnarray}
      \int_0^1 dx xG(x,Q^2\to\infty) & = & \cfrac{8}{17}, \nonumber \\
      \int_0^1 dx xq_s(x,Q^2\to\infty) & = & \cfrac{3}{68}\ \hbox{\rm
(each flavor),} \\
      \int_0^1 dx xq_v(x,Q^2\to\infty) & = & 0, \nonumber
\end{eqnarray}
expected in QCD with six quark flavors and no light colored superpartners.
It is easy to verify that the momentum sum rule \eqn{eq:sr} is satisfied:
\beq
      \cfrac{8}{17} + \hbox{6 flavors}\cdot\hbox{2 (quarks+antiquarks)}
\cdot\cfrac{3}{68} = 1.
\eeq

\section{The Top Quark Must Exist}
Ever since the existence of the $b$-quark was inferred from the discovery of 
the $\Upsilon$ family of resonances in 1977,\cite{upsleon} we have been 
on the lookout for its weak-isospin partner, called top.  The long search, which  
occupied experimenters at laboratories around the world, came to a 
successful conclusion in 1995 with the announcement 
that the top quark had been observed in the CDF~\cite{CDF} and 
D\O~\cite{Dzero} experiments at Fermilab.\cite{WWrev}

Although top has now been established and is under close experimental scrutiny, 
it is worth reviewing the arguments that convinced us that top had to 
exist.  Even before we had direct experimental evidence for $b$ and 
$\tau$, M.~Kobayashi and T.\ Maskawa~\cite{KM} raised the possibility 
that \textsl{CP} violation arises from complex elements of the quark mass 
matrix, if there are at least three fermion generations.
Once the charge of the 
$b$-quark was established to be $e_b = -\cfrac{1}{3}$, it was natural to expect 
that the missing partner 
should be the upper member of a doublet, with charge $+\cfrac{2}{3}$.  
Completing the weak-isospin doublet by finding the top quark
is the most natural way to cancel the 
anomaly of the $(\nu_\tau, \tau)_L$ doublet and ensure an 
anomaly-free electroweak theory. 

The absence of flavor-changing neutral currents in $b$ decays, which 
would lead to significant branching fractions for dilepton channels such as 
$b\rightarrow s \ell^+ \ell^-$, added phenomenological support to the 
idea that $b$ is a member of a left-handed weak doublet.

More recently, the accumulation of results on the neutral-current 
interactions of the $b$-quark has made it possible to characterize the 
$Zb\bar{b}$ vertex and measure the weak isospin of the 
$b$-quark.  Consider a generalization of the $SU(2)_L\otimes U(1)_Y$ 
theory in which the $b$-quark may carry both left-handed and right-handed 
weak isospin.  The chiral neutral-current couplings can be written as
\begin{equation}
	\begin{array}{rcl}
		L_b & = & I_{L3} - e_b \sin^2\theta_W  \\
		
		R_b & = & I_{R3} - e_b \sin^2\theta_W\;\;, 
	\end{array}
	\label{chiNC}
\end{equation} which differ from the standard-model chiral couplings by 
the presence of $I_{R3}$.  Characteristics of the reaction $e^+ e^- 
\rightarrow b\bar{b}$ permit us to determine the values of $I_{L3}$ and 
$I_{R3}$ directly from experiment.  
The partial width $\Gamma(Z^0 \rightarrow b\bar{b})$ measures the 
combination $L_b^2 + R_b^2$.  On the $Z^0$ resonance, the 
forward-backward asymmetry $A_{{\rm FB}}(Z^0\rightarrow b\bar{b})$ 
measures the combination $(L_b^2 - R_b^2)/(L_b^2 + R_b^2)$.  Far below 
the resonance, where the forward-backward asymmetry is dominated by 
$\gamma$-$Z$ interference, $A_{{\rm FB}}(e^+ e^-\rightarrow b\bar{b})$ 
measures the combination $L_b - R_b$.  
The unique overlap of the allowed regions is for $I_{L3}=-\cfrac{1}{2}$, 
$I_{R3}=0$, the standard-model solution.
 
\section{Anticipating $m_t$}
Little can be said on general theoretical grounds about the masses of new 
flavors, but interesting constraints arise from consistency requirements 
and from phenomenological relationships.  Imposing the requirement that 
partial-wave unitarity be respected at the tree level in the reactions
\begin{equation}
	Q\bar{Q}\rightarrow (W^+W^-,Z^0Z^0,HZ^0,HH)
	\label{Qunit}
\end{equation}
leads to a condition on the heavy-quark mass $m_Q$, which determines the scale 
$(G_Fm_Q^2\sqrt{2})^{1/2}$ of the $HQ\bar{Q}$ couplings.\cite{Qmass}  
For the $(t,b)_L$ doublet of heavy quarks, the restriction amounts to
\begin{equation}
	|m_t - m_b| \ltap 550\gevcc.
	\label{tbound}
\end{equation}
This general constraint can be sharpened appreciably by considering 
radiative corrections to electroweak observables.

\begin{figure}[tb]
	\centerline{\BoxedEPSF{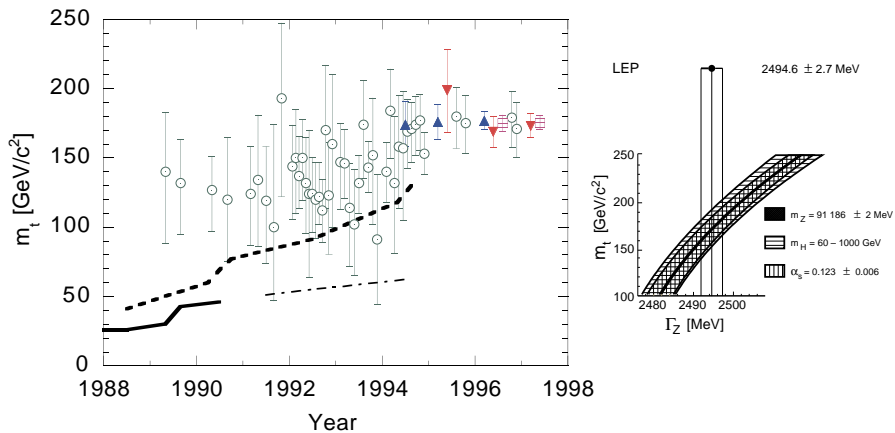  scaled 450}}
	\caption{Indirect determinations of the top-quark mass from fits to 
	electroweak observables (open circles) and 95\% confidence-level
	lower bounds on the top-quark 
	mass inferred from direct searches in $e^{+}e^{-}$ annihilations 
	(solid line) and in $\bar{p}p$ collisions, assuming that standard 
	decay modes dominate (broken line).  An indirect lower bound, derived 
	from the $W$-boson width inferred from $\bar{p}p \rightarrow 
	(W\hbox{ or }Z)+\hbox{ anything}$, is shown as the dot-dashed line.  
	Direct measurements of $m_{t}$ by the CDF (triangles) and D\O\ 
	(inverted triangles) Collaborations are shown at the time of initial evidence, 
	discovery claim, and today.  The current world 
	average from direct observations is shown as the crossed box.  For 
	sources of data, see Ref. {\protect\cite{pdg}}.  \textit{Inset:} Electroweak 
	theory predictions for the width of the $Z^{0}$ boson as a function 
	of the top-quark mass, compared with the width measured in LEP  
	 experiments (Ref. {\protect\cite{zwidth}}).}
	\protect\label{EWtop}
\end{figure}

Higher-order processes involving virtual top quarks are an important element 
in quantum 
corrections to the predictions the electroweak theory makes for many 
observables.  A case in point is the total decay rate, or width, of 
the $Z^{0}$ boson, which has been measured to exquisite precision at 
the CERN and SLAC $Z$ factories.  The comparison of experiment and 
theory shown in the inset to Figure \ref{EWtop} favors a top mass in 
the neighborhood of $180\gevcc$.  The top mass favored by simultaneous 
fits to many electroweak observables is shown as a function of time in 
Figure \ref{EWtop}.

\begin{figure}[t!]
	\centerline{\BoxedEPSF{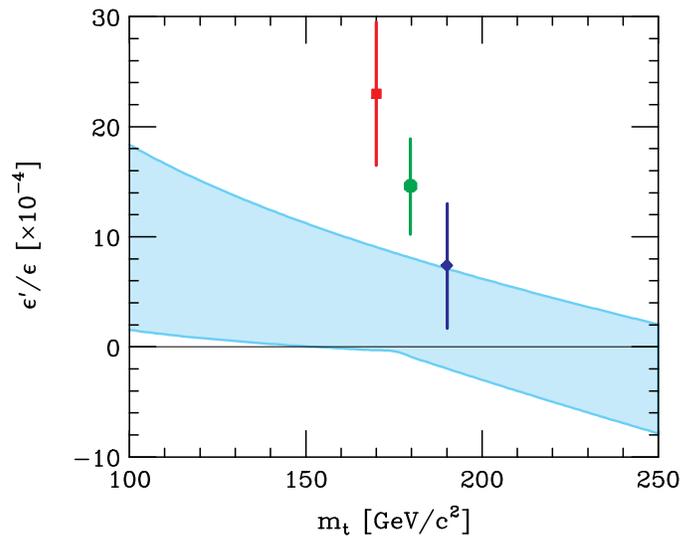  scaled 600}}
	\caption[epspr]{The quantity $\epsilon^\prime/\epsilon$ in the standard model 
	as a function of the top-quark mass.  The band shows the region allowed 
	by plausible variations in theoretical parameters.  The values measured 
	by NA31 (\protect\rule[-0.5pt]{4pt}{4pt}) and E731(\diamant) are shown, together with the 
	world average (\begin{picture}(6,10) 
	\put(3,2.5){\circle*{5}}\end{picture}).} 
\label{fig:gerhard}
\end{figure}

Many other observables, particularly those related to neutral-meson 
mixing and \textsl{CP} violation, are sensitive to the top-quark mass.  One 
example, for which we may expect significant progress over the next five 
years, is the parameter $\epsilon^\prime$ that measures direct \textsl{CP} 
violation in the $K^0$-$\bar{K}^0$ system.  Figure \ref{fig:gerhard} 
shows the region favored by state-of-the-art calculations as a function 
of $m_t$.\cite{buras}  We expect the theoretical uncertainty to shrink as lattice-QCD 
calculations mature.  The values measured by E731 at Fermilab and
by NA31 at CERN  are plotted at arbitrary values 
of $m_t$.\cite{epspr}  The new generation of experiments may reduce the 
experimental uncertainty on $\epsilon^\prime/\epsilon$ to $\pm 1 \times 10^{-4}$.

It is worth mentioning another hint that I have to confess seems more 
suggestive to me after the fact than it did before.  In 
supersymmetric unified theories of the fundamental interactions, 
virtual top quarks can drive the spontaneous breakdown of 
electroweak symmetry---provided top is very massive.\cite{AGPW}

\section{Production Rates \label{pr}}
\begin{figure}[tb]
	\centerline{\BoxedEPSF{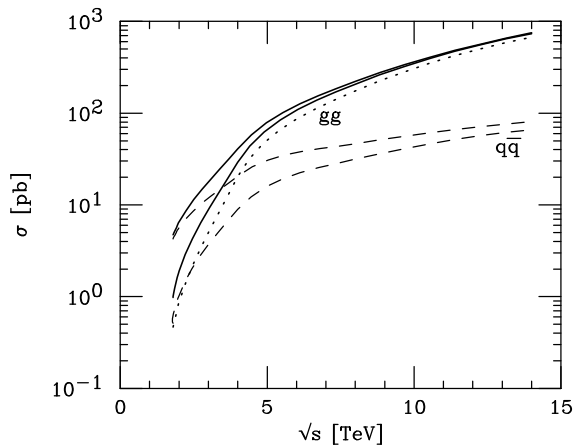  scaled 500}}
	\caption{Energy dependence of the cross section for production of 
	175-GeV$\!/\!c^2$ top quarks
	in $pp$ (lower curves) and $\bar{p}p$ (upper curves) collisions.  The 
	contributions of $\bar{q}q$ (dashed curves) and $gg$ (dotted curve) 
	collisions are  
	shown separately. (After Parke, Ref. {\protect\cite{sparkieABQ}}.)} 
\label{fig:sigtop}
\end{figure}

The calculation of the top-quark production cross section in perturbative 
QCD has been carried out to next-to-leading order (NLO) and beyond, using 
resummation techniques.\cite{laenen}  Typical results are shown in 
Figure \ref{fig:sigtop} for $p^\pm p$ collisions at c.m.\ energies 
between 1.8 and $14\tev$.    At $1.8\tev$, for 
$m_t=175\gevcc$, the QCD cross section is 
$\sigma(\bar{p}p\rightarrow t\bar{t}+\hbox{ anything}) \approx 6\pb$, 
predominantly from 
the elementary process $\bar{q}q \rightarrow t\bar{t}$.  
At the level of $\pm30$\%, there are differences among the competing 
calculations that need to be resolved.
  At 14\tev, the 
energy planned for the Large Hadron Collider at CERN, the QCD cross 
section rises to 
$\sigma(pp\rightarrow t\bar{t}+\hbox{ anything}) \approx 800\pb$, 
predominantly from the mechanism $gg \rightarrow t\bar{t}$.

\begin{figure}[tb]
	\centerline{\BoxedEPSF{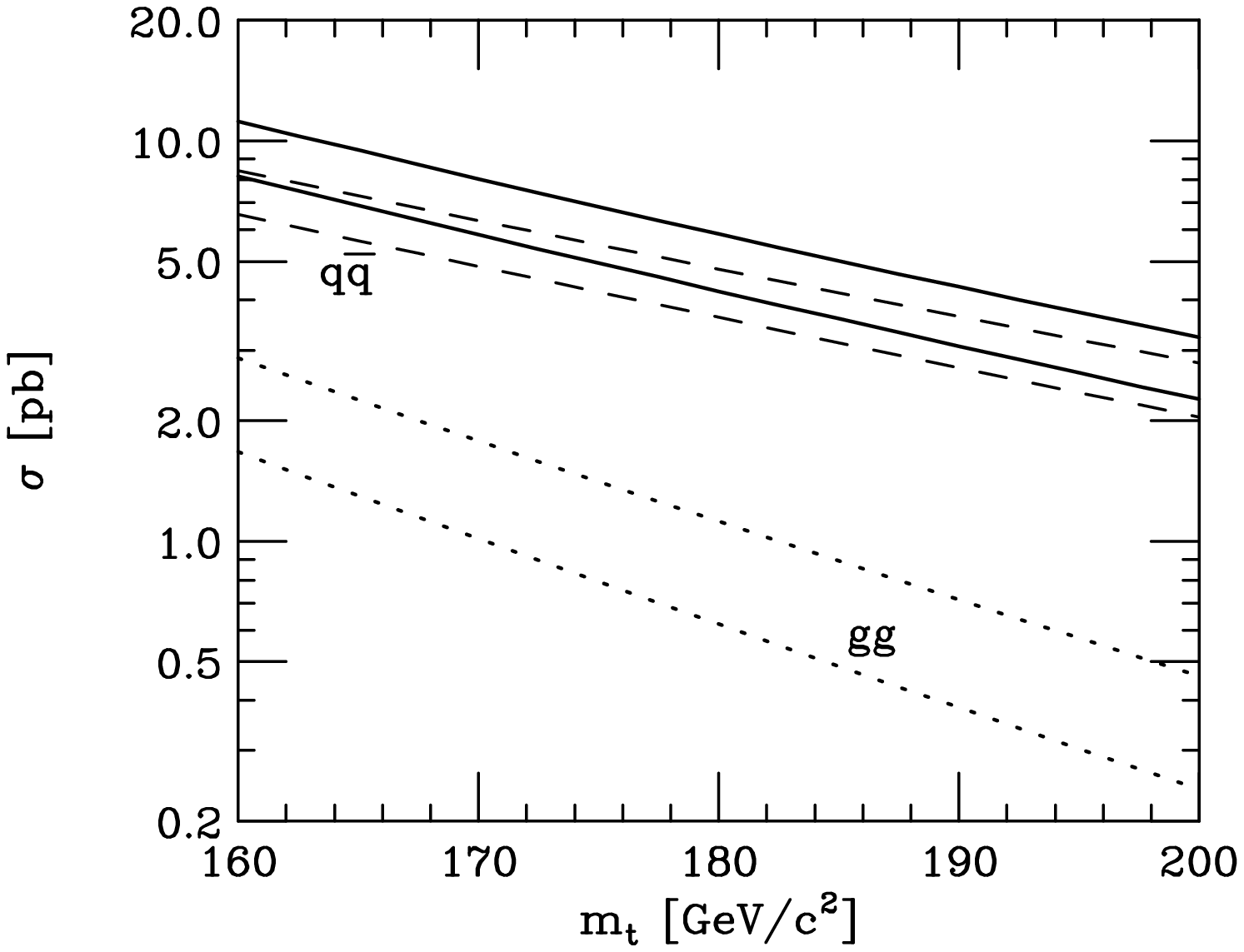  scaled 500}}
	\caption{Dependence of the top production cross section, 
	$\sigma(\bar{p}p\rightarrow t\bar{t}+\hbox{ anything})$, upon the 
	top-quark mass in 1.8-TeV (lower curves) and 2.0-TeV (upper curves) 
	$\bar{p}p$ collisions.  The 
	contributions of $q\bar{q}$ (dashed curves) and $gg$ (dotted curves) are 
	shown separately.} \label{fig:Tevtop}
\end{figure}

It is interesting to ask what would be gained by raising the top energy 
of the Tevatron collider by lowering the operating temperature of the 
superconducting magnets.  Figure \ref{fig:Tevtop} shows how the top 
production cross section depends on $m_t$ at 
$\sqrt{s}=1.8\hbox{ and }2.0\tev$, according to the resummed 
next-to-leading-order calculation of Laenen, Smith, and van 
Neerven.\cite{eric}  For 
$160\gevcc \le m_t \le 200\gevcc$, the cross section will increase by about 
40\% when the c.m.\ energy is raised to 2\tev.  The fraction of the cross 
section contributed by $gg$ collisions grows from about 15\% to about 
20\% for a top-quark mass of $175\gevcc$.
\begin{figure}[tb]
	\centerline{\BoxedEPSF{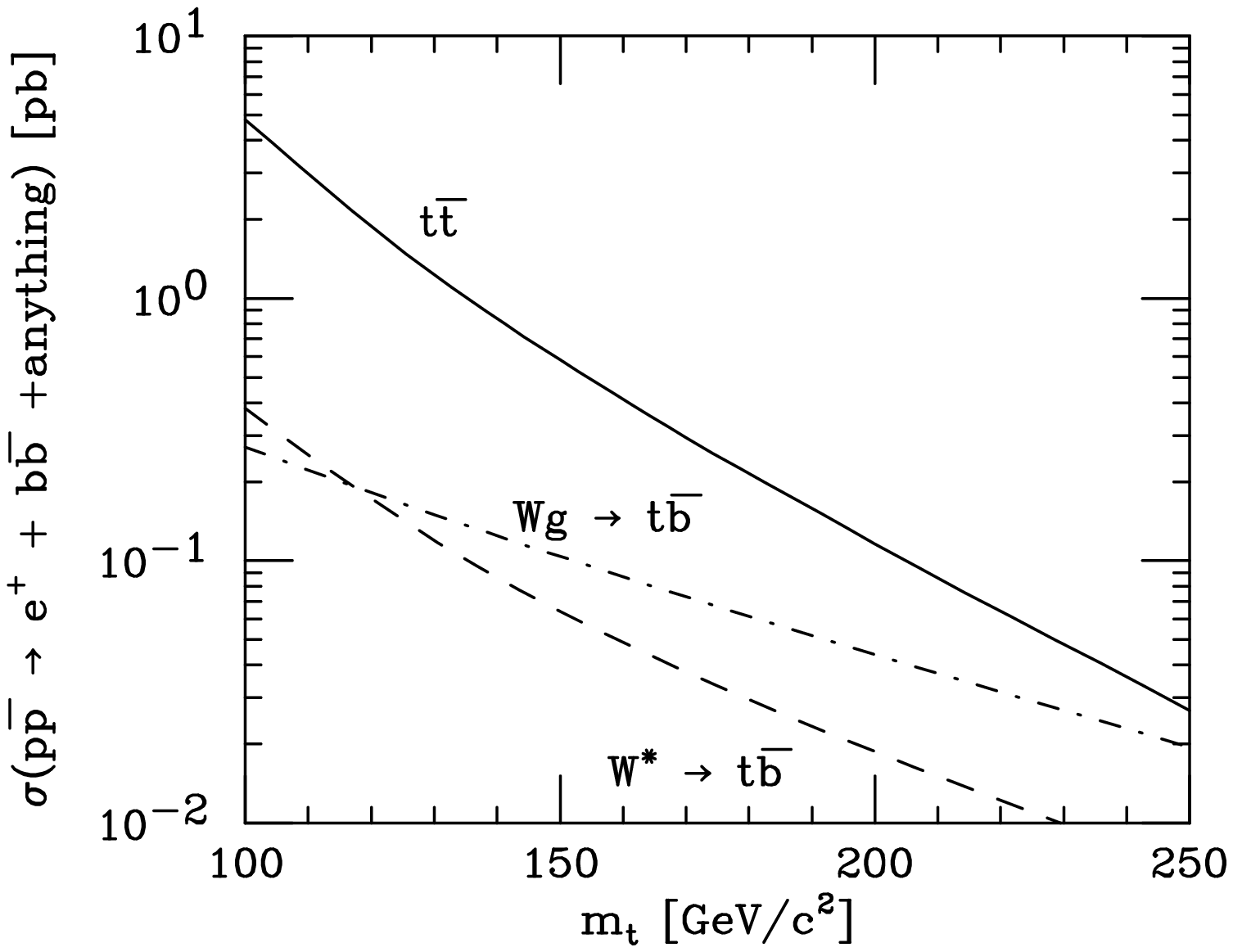  scaled 500}}
	\caption[dum]{Variation of standard-model contributions to the cross section 
	$\sigma(\bar{p}p \rightarrow e^+ b\bar{b}+\hbox{ jets})$ with the 
	top-quark mass at $\sqrt{s}=1.8\tev$.  Yields are shown for the 
	$t\bar{t}$ (solid 
	curve), $Wg \rightarrow t\bar{b}$ (dot-dashed curve), and $W^\star 
	\rightarrow t\bar{b}$ (dashed curve) contributions.  [After Parke, Ref. 
	{\protect\cite{sparkieABQ}}.]} \label{fig:othertop}
\end{figure}
In addition to the dominant mechanisms for top production included in 
Figures \ref{fig:sigtop} and \ref{fig:Tevtop}, other conventional sources may take on 
importance as the integrated luminosity rises.  I show in Figure 
\ref{fig:othertop} the contributions of the weak-interaction processes 
$W^+g \rightarrow t\bar{b}$ and (virtual) $q\bar{q} \rightarrow W^\star 
\rightarrow t\bar{b}$ to the cross section for producing $e^+ +\hbox{ jets}$,
assuming that $t\rightarrow bW^+$ is top's only decay mode.  
The final state contains $e^+ b \bar{b}+ (2,1,0)\hbox{ non-$b$-quark 
jets}$ for the QCD, $W$-gluon, and virtual-$W$ processes.  

\section{Top Width and Lifetime}
In the standard model, the dominant decay of a heavy top quark is the 
semiweak process $t \rightarrow bW^+$, for which the decay rate is \cite{Gamt}
\begin{eqnarray}
	\Gamma(t \rightarrow bW^+) & = & 
	\frac{G_FM_W^2}{8\pi\sqrt{2}}\frac{1}{m_t^3}
	\left[\frac{(m_t^2-m_b^2)^2}{M_W^2}+m_t^2+m_b^2-2M_W^2\right] \nonumber\\
	 & & \times	\sqrt{[m_t^2-(M_W+m_b)^2][m_t^2-(M_W-m_b)^2]}.
	\label{tdk}
\end{eqnarray}
Here $m_{t}$, $m_{b}$, and $M_{W}$ are the masses of top, bottom, and 
the $W$-boson, and $V_{tb}$ measures the strength of the $t 
\rightarrow bW^{+}$ coupling.  To the extent that the $b$-quark mass 
is negligible, 
the decay rate can be recast in the form
\begin{displaymath}
	\Gamma(t \rightarrow bW^+) = \frac{G_F m_t^3}{8\pi\sqrt{2}}
	|V_{tb}|^{2} \left(1 - \frac{M_{W}^{2}}{m_{t}^{2}}\right)^{\!\!2}
	\left(1 + \frac{2M_{W}^{2}}{m_{t}^{2}}\right) ,
	\label{tdkapp}
\end{displaymath}
which grows rapidly with increasing top mass.

If there are only three generations of quarks, so that the 
Cabibbo-Kobayashi-Maskawa matrix element $V_{tb}$ has a magnitude close 
to unity, then for a top-quark mass of $175\gevcc$, the partial width is
\begin{equation}
	\Gamma(t \rightarrow bW^+) \approx 1.55\gev,
	\label{eqn:twidth}
\end{equation} which corresponds to a top lifetime $\tau_t \approx 0.4 
\times 10^{-24}\hbox{ s}$, or 0.4 yoctosecond (ys).\cite{SI}  
The confining effects of the strong interaction act on a time scale 
of a few yoctoseconds set by $1/\hbox{the scale energy of quantum 
chromodynamics}$, $\Lambda_{\mathrm{QCD}}$.
This means that a top quark decays long before it can be hadronized.
There will be no discrete lines in toponium ($t\bar{t}$) spectroscopy, and indeed 
no dressed hadronic states containing top.  
Accordingly, the characteristics of top 
production and the hadronic environment near top in phase space should be 
calculable in perturbative QCD.\cite{pqcd}

\begin{figure}[tb]
	\centerline{\BoxedEPSF{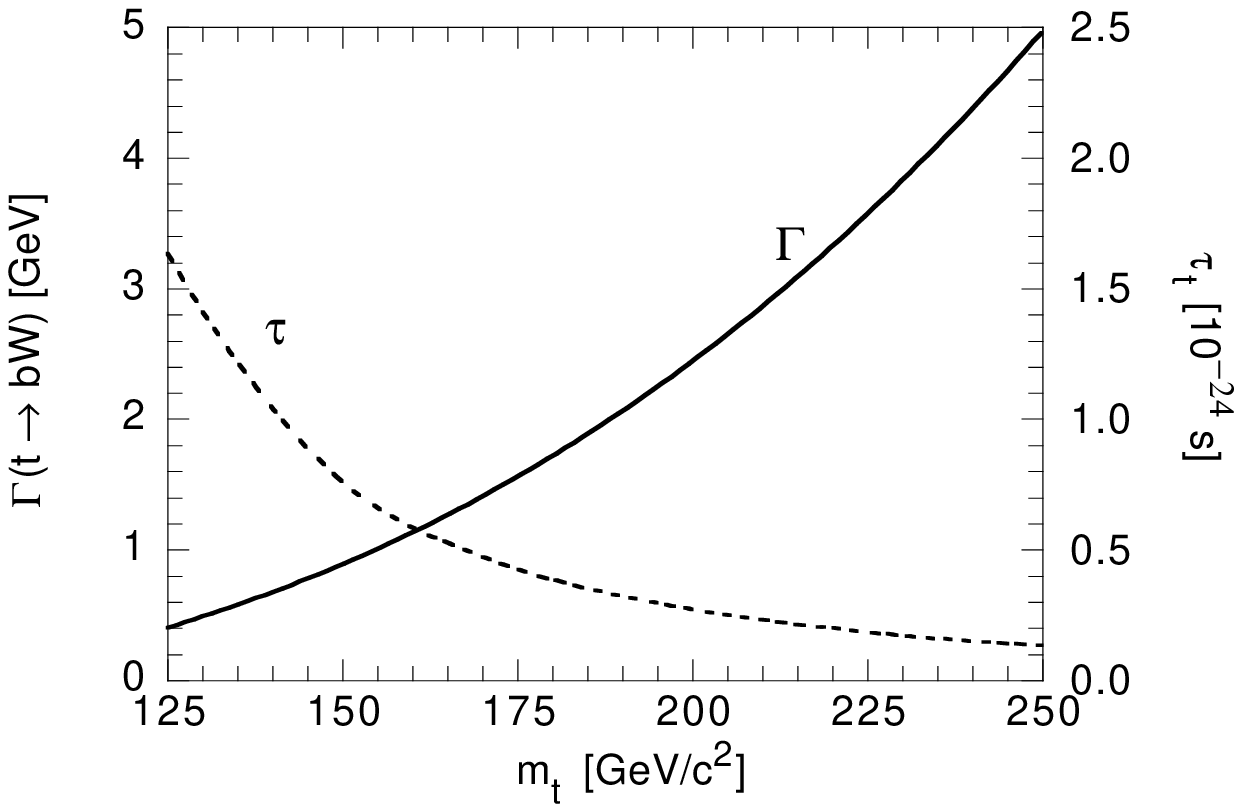  scaled 750}}
	\caption{Partial width (solid curve, left-hand scale) for the decay 
	$t \rightarrow bW^+$ as a function of $m_t$.  The (partial) lifetime 
	 is shown as the dashed curve (right-hand scale).  A full-strength 
	 $t\rightarrow b$ transition is assumed, and QCD corrections are 
	 omitted.}
	\protect\label{fig:topwidth}
\end{figure}

The $t \rightarrow bW^+$ decay rate and (partial) lifetime are shown in 
Figure \ref{fig:topwidth} for a range of top-quark masses.

It is noteworthy that top decay is an excellent source of longitudinally 
polarized $W$-bosons, which may be particularly sensitive to new 
physics.  $W$-bosons with helicity $=-1$ are emitted with relative weight 1 and 
those with helicity $=0$ are produced with relative weight $m_t^2/2M_W^2$.  For 
$m_t = 180\gevcc$, a fraction $f_0 = 71$\% of the $W$-bosons 
emitted in top decay will be longitudinally polarized.  The decay angular 
distribution of charged leptons in the $W$ rest frame is
\begin{equation}
	\frac{d\Gamma(W^+\rightarrow\ell^+\nu_\ell)}{d(\cos\theta)} =
	\cfrac{3}{8}(1-f_0)(1-\cos\theta)^2 + \cfrac{3}{4}f_0 \sin^2\theta .
	\label{angular}
\end{equation}
\section{Top Search and Discovery}
Through the 1980s and early 1990s, direct searches continually raised the 
lower bound on the 
top mass, but produced no convincing sign of the top quark.  The most 
stringent limits (Cf. Figure \ref{EWtop}) came from the proton-antiproton colliders at CERN 
and Fermilab, but these 
relied on the assumption that top decays (almost) exclusively into a 
bottom quark and a real or virtual $W$ boson.  Electron-positron 
colliders could look for $e^{+}e^{-}\rightarrow t\bar{t}$ without 
assumptions about the decay mechanism, but the lower energies of 
those machines led to rather weak bounds on $m_{t}$.

By 1994, an impressive body of circumstantial evidence pointed to
the existence of a top quark with a 
mass of $175 \pm 25\gevcc$.  
Finding top and measuring its mass directly emerged as a critical test 
of the understanding of weak and electromagnetic interactions
built up over two decades.

The decisive experiments were carried out at Fermilab's Tevatron, in 
which a beam of 900-GeV protons collides with a beam of 900-GeV 
antiprotons.
Creating top-antitop pairs in sufficient numbers to claim discovery 
demanded 
exceptional performance from the Tevatron, for only one interaction in 
ten billion results in a top-antitop pair.  Observing traces 
of the disintegration of top into a $b$-quark and a $W$-boson
 required highly capable detectors and 
extraordinary attention to experimental detail.  Both the $b$-quark 
and the $W$-boson are themselves unstable, with many multibody decay 
modes.  The $b$-quark's mean lifetime is about $1.5\ps$.  It can be 
identified by a decay vertex displaced by a fraction of a millimeter 
from the production point, or by the low-momentum electron or muon from the 
semileptonic decays $b \rightarrow ce\nu$, $b \rightarrow c\mu\nu$, 
each with branching fraction about 10\%.  The $W$ boson decays after 
only $0.3\ys$ on average into $e\bar{\nu}_{e}$, 
$\mu\bar{\nu}_{\mu}$, $\tau\bar{\nu}_{\tau}$, or a quark and 
antiquark (observed as two jets of hadrons), 
with probabilities $\cfrac{1}{9}$, $\cfrac{1}{9}$, $\cfrac{1}{9}$, and 
$\cfrac{2}{3}$.  

The first evidence for top was presented in April 1994 by the CDF 
Col\-lab\-or\-a\-tion.\cite{cdf1}
In a sample of 19.3 events per picobarn of cross section 
($19.3\pb^{-1}$), CDF found 12 events consistent with 
either two $W$ bosons, or a $W$ boson and at least one $b$-quark.  
Although the sample lacked the 
statistical weight needed to claim discovery, the event 
characteristics were consistent with the $t\bar{t}$ interpretation, 
with a top mass of $174 \pm 10 ^{+13}_{-12}\gevcc$.  A few months 
later, the D\O\ Collaboration reported an excess of candidates (9 
events with an expected background of $3.8 \pm 0.9$) in a 
13.5-pb$^{-1}$ sample.\cite{dzero1}

The discovery was not far behind.  By February 1995, both groups had 
quadrupled their data sets.  The CDF Collaboration found 6 
dilepton candidates with an anticipated background of $1.3 \pm 0.3$ 
events, plus 37 $b$-tagged events containing a $W$-boson and at least 
three jets.\cite{CDF}  The D\O\ 
Collaboration reported 17 top candidates with an expected 
background of $3.8 \pm 0.6$.\cite{Dzero}  Taken 
together, the populations and characteristics of different event 
classes provided irresistible evidence for a top quark with a mass in 
the anticipated region: $176\pm 8 \pm 10\gevcc$ for 
CDF, and $199^{+19}_{-21}\pm 22\gevcc$ for D\O.  The 
top-antitop production rate was roughly in line with theoretical predictions.

Today, with the event samples approximately doubled again, the top 
mass is measured as $176.8 \pm 6.5\gevcc$ by CDF and $173.3 \pm 
8.4\gevcc$ by D\O\, for a world average of $175.5 \pm 
5.1\gevcc$.\cite{97vals}  The production cross sections, 
$\sigma(\bar{p}p \rightarrow t\bar{t}+\hbox{anything}) = 5.53 \pm 
1.67\pb$ (D\O\ ) and $7.5^{1.9}_{1.6}\pb$ (CDF) are close to what is 
expected.

The observation of top completes the last normal (light-neutrino) 
generation and provides a crucial parameter of the electroweak theory.  
The large mass of the top quark suggests that top might stand apart from 
the other quarks and leptons.  Top  provides a new window on novel 
physics through nonstandard production and decay.  We now take up some of 
the implications of the discovery.

\section{Top in Electroweak Radiative Corrections}
The influence of the top quark on electroweak radiative corrections was 
the basis for the expectations for $m_t$ from precision measurements of 
electroweak observables.  As the top-quark mass is known more precisely 
from direct measurements, it will be possible to compare predictions for 
which $m_t$ is an input with other observations.  Over the next few 
years, we can anticipate incisive tests of the electroweak theory from 
the comparison of the $W$-boson mass with theoretical calculations.

\begin{figure}[tbp]
	\centerline{\BoxedEPSF{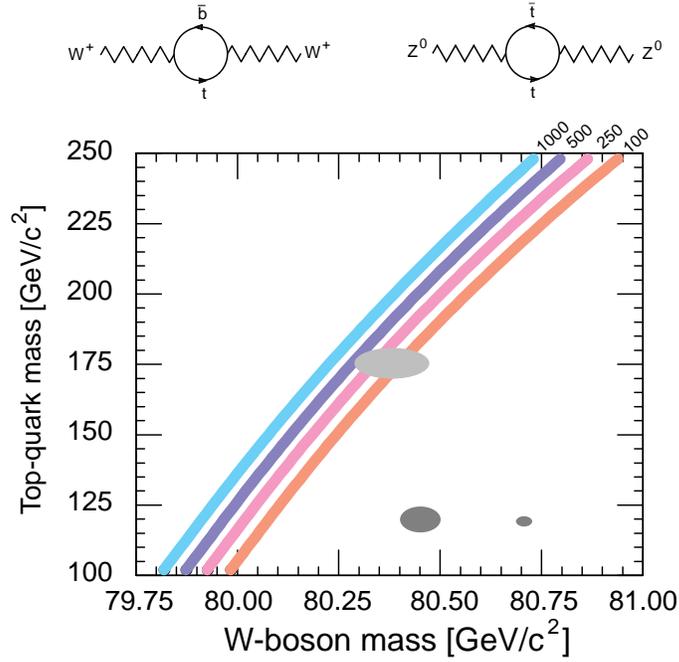  scaled 575}}
	\caption{Correlation between the top-quark mass and the $W$-boson 
	mass in the standard electroweak theory.  From left to 
	right, the bands correspond to Higgs-boson masses of $1000,\:500,\: 
	250, \hbox{ and }100\gevcc$.  The thickness of the bands expresses 
	the effect of plausible variations in the value of 
	$\alpha(M_{Z})$.  The dark region is the 
	one-standard-deviation error ellipse from the current world 
	averages, $m_{t}=175.5 \pm 5.1\gevcc$ and $M_{W} = 80.38 \pm 
	0.09\gevcc$.  
	Also shown are the one-standard-deviation error ellipses for 
	precisions expected in the future: ($\delta M_{W}=50\mevcc, \;\delta 
	m_{t}=5\gevcc$) and ($\delta M_{W}=20\mevcc,\; \delta 
	m_{t}=2\gevcc$).  Examples of the heavy-quark loops that give rise to 
	$\Delta\rho$ are shown at the top of the figure.}
	\protect\label{fig:mwmt}
\end{figure}

The $W$-boson mass is given as
\begin{equation}
	M_{W}^{2} = M_{Z}^{2} (1 - \sin^{2}\theta_{W})(1 + \Delta\rho),
	\label{rho}
\end{equation}
where $M_{Z}$ is the mass of the $Z^{0}$ boson, 
$\sin^{2}\theta_{W}\approx 0.232$ is the weak mixing parameter, and 
$\Delta\rho$ represents quantum corrections.  Some of the most important of 
these are shown at the top of Figure \ref{fig:mwmt}.  The inequality of the $t$- 
and $b$-quark masses violates weak-isospin symmetry and results in
\begin{equation}
	\Delta \rho = 3G_{F}m_{t}^{2}/8\pi^{2}\sqrt{2} + \ldots,
	\label{deltarho}
\end{equation}
where the unwritten terms include a logarithmic dependence upon the 
mass of the Higgs boson, the hitherto undetected agent of electroweak 
symmetry breaking.

Predictions for $M_{W}$ as a function of the top-quark mass are shown 
in Figure \ref{fig:mwmt}
for several values of the Higgs-boson mass.\cite{tatsu}  Current 
measurements are consistent with the electroweak theory, but do not 
yet provide any precise hints about the mass of the Higgs boson.  The 
uncertainty on the world-average $M_{W}$ has now reached about $100\mevcc$.  
An uncertainty of $\delta M_{W}=50\mevcc$ seems a realistic 
possibility both at the Tevatron and at CERN's LEP200, where 
observations of the reaction $e^{+}e^{-}\rightarrow W^{+}W^{-}$ near 
threshold began in 1996.  Improving $\delta m_{t}$ below $5\gevcc$ will 
then make for a demanding test of the electroweak theory that should 
yield interesting clues about the Higgs-boson mass.  Over the next decade, 
it seems 
possible to reduce $\delta m_{t}$ to $2\gevcc$ at Fermilab and $\delta 
M_{W}$ to about $20\mevcc$ at the Tevatron and LEP200.  That will set 
the stage for a crucial test of the electroweak theory when (and if) the Higgs 
boson is discovered.

\section{Is It Standard Top?}
The top-quark discovery channels all arise from the production of 
top-antitop pairs.  We expect that all the significant channels will 
contain a $b\bar{b}$ pair, from the decay chain
	\twodk{t}{\bar{t}}{bW^+\;,}{\bar{b}W^-}
leading to $b\bar{b}e^\pm \mu^\mp\nu\nu$, $b\bar{b}e^+ 
e^-\nu\nu$, $b\bar{b}\mu^+\mu^-\nu\nu$, $b\bar{b}\ell\nu\hbox{ jet jet}$.

We expect that decays other than the observed $t \rightarrow bW^+$ mode 
are strongly suppressed.  Unless the quark--mixing-matrix element 
$\abs{V_{tb}} \ll 1$, which could occur if 
top had a strong coupling to a fourth-generation $b^\prime$ with 
$m_{b^\prime} > m_t$, the decays $t \rightarrow (s,d)W^+$ should be 
extremely rare.\cite{howie}
It is important to test this expectation by looking for 
the rare decays directly, or by comparing the number of observed 
$(0,1,\hbox{ and }2)$ $b$-tags in a top-quark sample with expectations 
derived from the various top-production mechanisms and the efficiency for 
$b$-tagging.  The CDF 
Collaboration has used the tagging method to show that $t \rightarrow 
bW$ accounts for $99\pm 29\%$ of all $t \rightarrow 
W+\hbox{ anything}$ decays.\cite{joei} 

Stelzer and Willenbrock have argued recently that the 
$W^\star \rightarrow t\bar{b}$ process may in time provide the best 
measurement of the quark mixing-matrix element $\abs{V_{tb}}$.\cite{sw}  
Prospects for extracting top-quark (and other) parameters from threshold 
studies at a future $e^+e^-$ linear collider have been surveyed by Fujii, 
Matsui, and Sumino.\cite{nlc} 

The rapid decay of the top quark means that there is no time for the 
formation of top mesons or baryons.  Accordingly, the spin orientation of 
the top quark at the moment of its production is reflected, without 
dilution, in the decay angular distribution of its decay products.  The 
lepton angular distribution thus becomes a tool for probing the structure 
of the charged-current interactions of top.\cite{spin}

The persistence of the top quark's polarization can be exploited to 
devise tests for \textsl{CP} violation in top decays.  Because the 
standard model leads to very tiny effects, the top system has great 
sensitivity to nonstandard sources of \textsl{CP} violation.  A brief 
review of the considerable literature on the subject can be found in Ref. 
\cite{cpv}. 

The branching ratios expected for the flavor-changing neutral-current decays
 \begin{equation}
 	t \rightarrow 
 	\left(\begin{array}{c}
 		g  \\
 		
 		Z  \\
 		
 		\gamma
 	\end{array}\right) + 
 	\left(\begin{array}{c}
 		c  \\
 		
 		u
 	\end{array}\right)
 	\label{eq:fcnc}
 \end{equation} all are unobservably small ($\ll 10^{-10}$) according to 
 the standard electroweak theory.\cite{fcnc}  Anomalous $Zt\bar{c}$ 
 couplings could  
 lead to a branching fraction as large as a few per cent while respecting 
 current constraints from low-energy phenomenology.  High-luminosity 
 experiments at the Tevatron or the LHC should be able to explore 
 branching fractions as small as $\sim10^{-2}$ and $\sim10^{-4}$, 
 re\-spec\-tive\-ly.\cite{anomfcnc}

Mahlon and Parke have examined the possibility that a 
threshold enhancement might render observable rare decays like 
$t\rightarrow bWZ$ and $t\rightarrow bWH$.\cite{rarissime}  The finite 
widths of the $W$ and $Z$ bosons do raise the decay rates dramatically 
near threshold, but the branching fractions remain too small to be 
observed in the current round of experiments.  Over the interval 
$160\gevcc \ltap m_t \ltap 200\gevcc$, the branching fraction 
$\Gamma(t\rightarrow bWZ)/\Gamma(t\rightarrow bW)$ rises from 
$1.6\times 10^{-7}$ to $1.4\times 10^{-5}$.  Detection of this mode in a 
modest top sample would therefore be a compelling sign of new physics.

Because top is so massive, many decay channels may be open to it, beyond 
the dominant $t \rightarrow bW^+$ mode.  The semiweak decay 
$t\rightarrow bP^+$, where $P^+$ is a charged (pseudo)scalar, may occur 
in multi-Higgs generalizations of the standard model, in supersymmetric 
models, and in technicolor models.  The decay rate $\Gamma(t\rightarrow 
bP^+)$ is generically comparable with $\Gamma(t\rightarrow bW^+)$, 
because both are semiweak.\cite{toplite}  An inferred $t\bar{t}$ 
production cross section smaller than that predicted by QCD would be a 
hint that $\Gamma(t\rightarrow bW^+)/\Gamma(t\rightarrow \hbox{all}) < 1$, 
which would argue for the presence of nonstandard decays.

Decay of a charged scalar into fermion pairs would typically proceed at a rate
\begin{equation}
	\Gamma(P^+ \rightarrow f_i\bar{f}_j) = \frac{G_F p(m_i^2+m_j^2)}{16\pi} 
	C_{ij}, 
	\label{Pdk}
\end{equation} where $p$ is the momentum of the products in the rest 
frame of $P^+$ and $C_{ij} = (3,1)$ for (quarks, leptons).  The lifetime 
of $P^+$ is far too short for it to be observed as a short track: for 
$M_{P^+} = M_W$, $\tau_{P^+} \ltap 10^{-21}\hbox{ s}$ (= 1~zeptosecond).\cite{Pmass}  
$P^+$ might be reconstructed from its decays into $c\bar{b}$ or 
$c\bar{s}$, or its presence might be deduced from $P^+ \rightarrow 
\tau^+\nu_\tau$ decays, which would also lead to violations of lepton 
universality.  

The general lesson is that top decays have the potential to surprise.  It 
may therefore be quite rewarding to learn to tag top-bearing events with 
high efficiency.\cite{CDFPsearch}

\section{The Dead Cone}
The large mass of the top quark has an important effect on the pattern of 
soft-gluon emission from an energetic top.  If the energy $\omega$ of the 
gluon is small compared to the energy $E_Q$ of the heavy quark, 
$\omega\ll E_Q$, then for a gluon emitted at a small angle $\theta \ll 1$ 
to the top-quark direction, the angular distribution of the radiation 
will be 
\begin{equation}
	d\sigma_{Q\rightarrow Qg} \sim \frac{\theta^2 
	d\theta^2}{(\theta^2+\theta_0^2)^2} \frac{d\omega}{\omega},
	\label{angrad}
\end{equation} where $\theta_0 = m_Q/E_Q$.  For angles larger than the 
critical value, \ie, for $\theta > \theta_0$, the radiation pattern becomes
\begin{equation}
	d\sigma \sim \frac{d\theta^2}{\theta^2} \frac{d\omega}{\omega} 
	\rightarrow d(\ln\theta^2)d(\ln\omega),
	\label{doublog}
\end{equation}which is doubly logarithmic.  Indeed, when 
$\theta\gg\theta_0$, the emission of successive gluons follows a strict 
angular ordering, and the multiplicity of hadrons accompanying the heavy 
quark is the same as it would be for a light quark.  In contrast, in the very 
forward cone defined by $\theta < \theta_0$, there is only a single 
logarithmic factor, $d\omega/\omega$.  In this region, gluon emission is 
inhibited and the multiplicity of accompanying hadrons is 
diminished.  The regime of reduced multiplicity is known as the dead 
cone.\cite{deadcone}  

Although the angular dependence of radiation accompanying a heavy quark has 
not been measured, there is some evidence for a reduced multiplicity in 
the number of hadrons emitted by an energetic $b$-quark.  The SLD and 
OPAL experiments have used tagged samples of $e^+e^- \rightarrow Z^0 
\rightarrow b\bar{b}$ events to compare the charged multiplicity 
$\langle n_b \rangle$ of 
hadrons produced by energetic $b$-quarks with the multiplicity 
$\langle n_{u,d,s} \rangle$ produced 
by energetic light quarks.\cite{dcexp}  SLD finds $ \langle n_{u,d,s} 
\rangle - \langle n_b \rangle
 = 3.31 \pm 0.41 \pm 0.79$, while OPAL measures 
$\langle n_{u,d,s} \rangle - \langle n_b \rangle = 3.02 \pm 0.05 \pm 
0.79$.  This is a significant suppression of particle emission, and bodes 
well for the possibility of reconstructing self-tagging $B^{**} 
\rightarrow B^{(*)}\pi$ decays cleanly for studies of \textsl{CP} 
violation in $B$ decays.  The effect should be 
considerably larger for top quarks, which will have the added advantage 
of decaying before they can be dressed into resonances whose effects are 
not included in the perturbative analysis.
  
\section{Top Matters}
It is popular to say that 
top quarks were produced in great numbers in the 
fiery cauldron of the Big Bang some fifteen billion years ago, disintegrated 
in the merest fraction of a 
second, and vanished from the scene until my colleagues learned to create 
them in the Tevatron.  That 
would be reason enough to care about top: to learn how it helped sow the 
seeds for the primordial universe that evolved into our world of diversity 
and change.  But it is not the whole story; it invests the top quark with a 
remoteness that veils its importance for the everyday world.

The real wonder is that here and now, every minute of every day, the top 
quark affects the world around us.  Through the uncertainty principle of 
quantum mechanics, top quarks and antiquarks wink in and out of an 
ephemeral presence in our world.  Though they appear virtually, fleetingly, 
on borrowed time, top quarks have real effects.

Quantum effects make the coupling strengths of the fundamental 
in\-ter\-ac\-tions---appropriately normalized analogues of the 
fine-structure constant $\alpha$---vary with the energy scale on 
which the coupling is measured.  The fine-structure constant itself 
has the familiar value $1/137$ in the low-energy (or long-wavelength) 
limit, but grows to about $1/129$ at the mass of the $Z^{0}$ boson, 
about $91\gevcc$.  Vacuum-polarization effects make the effective 
electric charge increase at short distances or high energies.

\begin{figure}[tb]
	\centerline{\BoxedEPSF{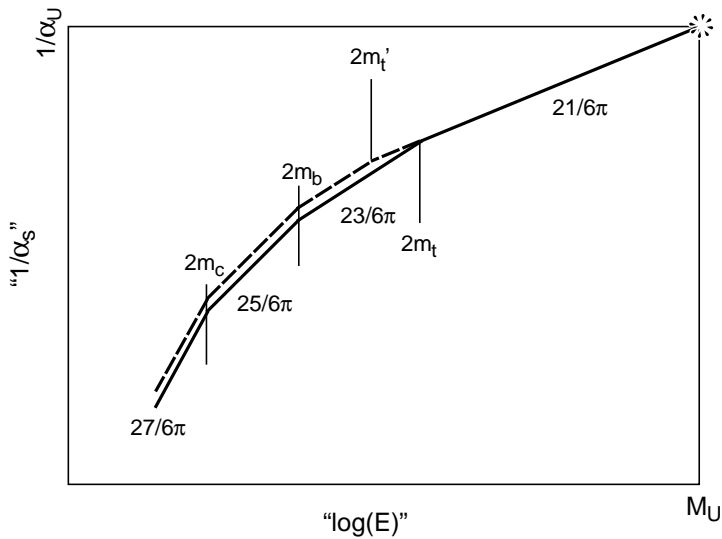  scaled 600}}
	\caption{Two evolutions of the strong coupling constant 
	$\alpha_{s}$.  A smaller value of the top-quark mass leads to a 
	smaller value of $\alpha_{s}$.}
	\protect\label{fig4}
\end{figure}

In unified theories of the strong, weak, and electromagnetic 
in\-ter\-ac\-tions, all the  coupling ``constants'' take on a 
common value, $\alpha_U$, at some high energy, $M_U$.  If we adopt the 
point of view that
$\alpha_{U}$ is fixed at the unification 
scale, then the mass of the top quark is encoded in the value of 
the strong coupling $\alpha_s$ that we 
experience at low energies.\cite{su5}  Assuming three generations of quarks and 
leptons, we  evolve $\alpha_s$ downwards in energy from the 
unification scale in the spirit of Georgi, Quinn, and Weinberg.\cite{GQW}
The leading-logarithmic behavior is given by
\begin{equation}
1/\alpha_s(Q) = 1/\alpha_U + \cfrac{21}{6\pi}\ln(Q/M_U)\;\; ,
\end{equation} for $M_U > Q > 2 m_t$.  The positive coefficient 
$+21/6\pi$ means that the strong coupling constant 
$\alpha_{s}$ is smaller at high energies than at low energies.  This 
behavior---opposite to the familiar behavior of the electric 
charge---is the celebrated property of asymptotic freedom.
In the interval between $2m_t$ and $2m_b$, the 
slope $(33-2n_{\!f})/6\pi$ (where $n_{\!f}$ is the number of active quark 
flavors) steepens to $23/6\pi$, and then increases by 
another $2/6\pi$ at every quark threshold.  At the boundary $Q=Q_n$ 
between effective field theories with $n-1$ and $n$ active flavors, the 
coupling constants $\alpha_s^{(n-1)}(Q_n)$ and $\alpha_s^{(n)}(Q_n)$ must 
match.  This behavior is 
shown by the solid line in Figure \ref{fig4}.

The dotted line in Figure \ref{fig4} shows how the evolution of 
$1/\alpha_s$ changes if the top-quark mass is reduced.  A smaller top 
mass means a larger low-energy value of $1/\alpha_s$, so a smaller 
value of $\alpha_s$.  

Neglecting the tiny ``current-quark'' masses of the up and down 
quarks, the scale parameter $\Lambda_{\hbox{\footnotesize QCD}}$ is the 
only mass parameter in QCD.  It determines the scale of the 
confinement energy that is the dominant contribution to the proton mass. 
To a good first approximation, 
\begin{equation}
	M_{\hbox{{\footnotesize proton}}} \approx C \Lambda_{\hbox{{\footnotesize QCD}}},
	\label{lattice}
\end{equation}
where the constant of proportionality $C$ is calculable using 
techniques of lattice field theory.
 
To discover the dependence of $\Lambda_{\hbox{{\footnotesize QCD}}}$ upon the top-quark 
mass, we calculate $\alpha_s(2m_t)$ 
evolving up from low energies and down from the unification scale, and match:
\begin{equation}
1/\alpha_U  +  {\displaystyle \cfrac{21}{6\pi}}\ln(2m_t/M_U) =  
 1/\alpha_s(2m_c) - {\displaystyle \cfrac{25}{6\pi}}\ln(m_c/m_b) 
 -{\displaystyle \cfrac{23}{6\pi}}\ln(m_b/m_t)   .
 \end{equation} Identifying
 \begin{equation}
 1/\alpha_s(2m_c) \equiv {\displaystyle\cfrac{ 
 27}{6\pi}}\ln(2m_c/\Lambda_{\hbox{{\footnotesize QCD}}})\;  ,
\end{equation}  we find that 
\begin{equation}
	\Lambda_{\hbox{{\footnotesize QCD}}}=e^{\displaystyle -6\pi/27\alpha_U} 
	\left(\frac{M_U}{1 \gev}\right)^{\!21/27} 
	\left(\frac{2m_t\cdot 2m_b\cdot 2m_c}{1\gev^{3}}\right)^{\!2/27}\gev \;\; .
	\label{blank}
\end{equation}  

We conclude that, in a simple unified theory,
\begin{equation}
	\frac{M_{\hbox{{\footnotesize proton}}}}{1\gev} \propto 
	\left(\frac{m_t}{1\gev}\right)^{2/27} \;\; ,
	\label{amazing}
\end{equation}
so that, for example, a factor-of-ten decrease in the top-quark mass 
would result in a 20\% decrease in the proton mass.  
This is a wonderful result.  Now, we can't use it to
compute the mass of the top quark, 
because we don't know the values of $M_{U}$ and $\alpha_{U}$, and 
haven't yet calculated precisely the constant of proportionality 
between the proton mass and the QCD scale parameter.  Never mind!  The 
important lesson---no surprise to any twentieth-century physicist---is
that the microworld does determine the behavior 
of the quotidian.  We will fully understand 
the origin of one of the most important parameters in the everyday 
world---the mass of the proton---only by knowing the properties of the 
top quark.\cite{18params}  

Top matters!

\section{Top's Yukawa Coupling}
In the $SU(2)_L\otimes U(1)_Y$ electroweak theory, Higgs scalars give 
masses to the electroweak gauge bosons $W^\pm$ and $Z^0$, and also to the 
elementary fermions.  While the gauge-boson masses are determined in 
terms of the weak mixing parameter $\sin^2\theta_W$, each fermion mass is 
set by a distinct Yukawa coupling, as
\begin{equation}
	m_f = \frac{G_f v}{\sqrt{2}} ,
	\label{fermass}
\end{equation}
where $v=(G_F\sqrt{2})^{-1/2} \approx 246\gev$.  The Yukawa coupling of 
the electron is $G_e \approx 3 \times 10^{-6}$.

The top quark stands apart from the other fundamental constituents 
because its Yukawa coupling is very close to unity: $G_t \approx 1$.  Is 
top special?  Or is it the only normal fermion?

\begin{figure}[t!]
	\centerline{\BoxedEPSF{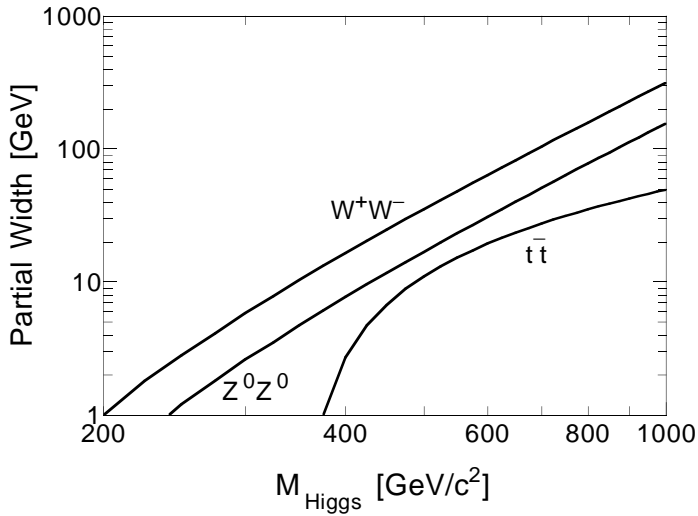  scaled 800}}
	\caption{Partial widths for the prominent decay modes of a heavy Higgs 
	boson.}
	\protect\label{fig:Hdk}
\end{figure}

In either case, the large $Ht\bar{t}$ coupling has several implications.  
(i) Higgs interactions will exert a significant influence on the 
evolution of the top-quark mass.  As we examine the possibility that the 
pattern of fermion masses is more intelligible at the unification scale 
than at the scale of common experience, it is important to evolve the 
fermion masses to our scale with care.\cite{schrempp2}  (ii) It is worth 
reexamining the  
reactions $q\bar{q} \rightarrow (\gamma^\star,Z^\star) \rightarrow t\bar{t}H$ and 
$q\bar{q} \rightarrow Z^\star \rightarrow ZH \rightarrow Zt\bar{t}$ at the 
LHC or, with $e^+e^-$ replacing $q\bar{q}$, at a multi-TeV linear 
collider.  (iii) A heavier top quark is a more important product of 
heavy-Higgs decay.  Figure \ref{fig:Hdk} shows the partial widths for 
Higgs-boson decay into the dominant $W^+W^-$ and $Z^0Z^0$ channels and 
into $t\bar{t}$, for $m_t = 175\gevcc$.  Whether the $t\bar{t}$ mode will 
be useful to confirm the observation of the Higgs boson, or merely drains 
probability from the favored $ZZ$ channel, is a question for detailed 
detector simulations.

\section{Are We Luckier Than We Deserve?}
According to the cockroach theory of stock market analysis, there is never 
a single piece of good news or bad news.  Might the discovery of top be 
the precursor of other discoveries?  More specifically, might the signal 
attributed to top also contain evidence of other new particles?  Let us 
review two possibilities.

A heavy top influences the spectrum analysis in minimal supersymmetric 
extensions of the standard model.  In some supersymmetric models, the 
lighter of the top squark eigenstates, called $\tilde{t}_1$, is less 
massive than the top quark: $m_{\tilde{t}_1} < m_t$.\cite{stop}  
Typically, the cross section for production of a heavy squark pair in 
$q\bar{q}$ interactions is about 1/8 to 1/4 of the cross section for 
production of a pair of quarks of the same mass.  Consequently, 
production of $\tilde{t}_1\tilde{t}_1^*$ is unlikely to distort the 
``top'' production cross section dramatically.  If it is kinematically 
allowed, the chain
\dkpp{\tilde{t}_1}{b}{\widetilde{W}^+}{W^+}{\tilde{\chi}^0}{\ell\nu\; ,}
where $\widetilde{W}^+$ is a wino and $\tilde{\chi}^0$ a neutralino, should 
be prominent among many decay channels.  It is a challenge to devise 
search strategies for the top squark and its decay products.

A second new quark, nearly degenerate with top, would have a stronger 
influence on ``$\sigma(t\bar{t})$.''  Barger and Phillips \cite{singletq} 
have explored the consequences of a weak-isoscalar, charge-2/3 quark, 
$t_s$, close in mass to top.  This singlet quark would decay by mixing 
with top ($t_s \rightarrow t \rightarrow bW^+$), so would populate the 
same decay modes.  By choosing $m_{t_s} \approx m_t$, it is easy to 
double the apparent top cross section.  Given the close agreement between 
measured and calculated cross sections for top production, we do not 
currently have a phenomenological incentive to do this.  However, the 
questions raised by this scenario are important and of general interest: 
Is the reconstructed top-quark mass distribution normal?  Is the 
$t\bar{t}$ effective-mass distribution normal?  Some numerical examples 
have been studied by Lane.\cite{weeken}

\section{Resonances in $t\bar{t}$ Production?}
Because objects associated with the breaking of electroweak symmetry tend 
to couple to fermion mass, the discovery of top opens a new window on 
electroweak symmetry breaking.  The possibility of new sources of 
$t\bar{t}$ pairs makes it urgent to test how closely top production 
conforms to standard (QCD) expectations.

Two classes of models have received considerable attention in the context 
of a heavy top quark.  Top-condensate models and multiscale 
technicolor both imply the existence of color-octet 
resonances that decay into $t\bar{t}$, for which the natural mass scale is 
a few hundred$\gevcc$.  We are led to ask:
Is there a resonance in $t\bar{t}$ production?
How is it made?  How (else) does it decay?

In the standard Higgs model of electroweak symmetry breaking, the Higgs 
potential breaks the $SU(2)_L\otimes U(1)_Y$ gauge symmetry.  The nonzero 
vacuum expectation value of the Higgs field endows the $W^\pm$ and $Z^0$ 
with mass and, through the arbitrary Yukawa couplings of the Higgs field 
to fermions, gives masses to the quarks and leptons.  The Higgs mechanism 
is a relativistic generalization of the Ginzburg--Landau phenomenology of 
the superconducting phase transition.  Both technicolor and topcolor are 
dynamical symmetry breaking schemes that are inspired by the 
Bardeen--Cooper--Schrieffer theory of superconductivity.

In technicolor, the QCD-like technicolor interaction becomes strong at 
low energies and forms a technifermion condensate that breaks chiral 
symmetry and gives masses to the gauge bosons $W^\pm$ and $Z^0$.  In a 
generalization of the basic scheme known as extended technicolor, new 
gauge bosons couple ordinary fermions to technifermions and allow the 
fermions to communicate with the technifermion condensate and acquire 
mass.\cite{techni}  In topcolor, a new interaction drives the formation 
of a $\langle\bar{t}t\rangle$ condensate that hides the electroweak 
symmetry and  
gives masses to the ordinary fermions.\cite{topcolor}

In the technicolor picture, which has been elaborated recently by Eichten 
and Lane,\cite{tc} a color-octet analogue of the $\eta^\prime$ meson, 
called $\eta_T$, is produced in gluon-gluon interactions.  The sequence
\begin{equation}
	gg \rightarrow \eta_T \rightarrow (gg, t\bar{t})
	\label{etaT}
\end{equation} leads to distortions of the $t\bar{t}$ invariant-mass 
distribution, and of the two-jet invariant-mass distribution.  Since 
$\Gamma(\eta_T \rightarrow b\bar{b})/\Gamma(\eta_T \rightarrow t\bar{t}) 
= m_b^2/m_t^2$, there should be only a negligible perturbation on the 
$b\bar{b}$ invariant-mass distribution.

In the topcolor picture explored by Hill and Parke,\cite{topsy} a 
massive vector ``coloron'' can be produced in $q\bar{q}$ interactions.  
The coloron decays at comparable rates into $t\bar{t}$ and $b\bar{b}$ and 
can appear as a resonance peak in both channels.  There is no clear 
reason to expect the coloron to distort the untagged two-jet 
invariant-mass spectrum.

If an enhancement were to be seen in the $t\bar{t}$ channel, a useful 
differential diagnostic, in addition to the $b\bar{b}$ and jet-jet 
invariant-mass distributions, will be the $t\bar{t}$ mass spectrum in 
different rapidity intervals and at different energies.  It is useful to 
recall from our discussion of top production rates in \S \ref{pr} that 
the relative $gg$ and $q\bar{q}$ luminosities at high masses change 
significantly between $\bar{p}p$ energies of 1.8 and 2.0\tev.  At the 
LHC, the large $gg$ luminosity would greatly enhance the contribution of 
$\eta_T$ with respect to the standard QCD process.  The 
isotropic decays of $\eta_T$ will help to characterize the technicolor 
case.  Hill and Parke \cite{topsy} and Lane \cite{weeken} have 
investigated the discriminatory power of $d\sigma/dp_\perp$ and of the 
production angular distributions for top.

When searching for resonances and other exotic sources of top, it is 
important not to apply cuts that efficiently exclude all mechanisms but 
standard QCD production.  Traditional expectations for sphericity, 
aplanarity, and similar event-shape parameters may not be realized for 
new sources. 

\section{Hiding a Gauge Symmetry}{\protect{\label{cache}}}
The most apt analogy for the hiding of the electroweak gauge 
symmetry is found in superconductivity.  In the Ginzburg-Landau 
description~\cite{4} of the superconducting phase transition, a 
superconducting material is regarded as a collection of two kinds of 
charge carriers: normal, resistive carriers, and superconducting, 
resistanceless carriers.

In the absence of a magnetic field, the free energy of the superconductor 
is related to the free energy in the normal state through
\begin{equation}
G_{\rm super}(0) = G_{\rm normal}(0) + \alpha \abs{\psi}^2 + \beta 
\abs{\phi}^4\;\;,
\end{equation}
where $\alpha$ and $\beta$ are phenomenological parameters and 
$\abs{\psi}^2$ is an order parameter corresponding to the density of 
superconducting charge carriers.  The parameter $\beta$ is non-negative, 
so that the free energy is bounded from below.

\begin{figure}
	\centerline{\BoxedEPSF{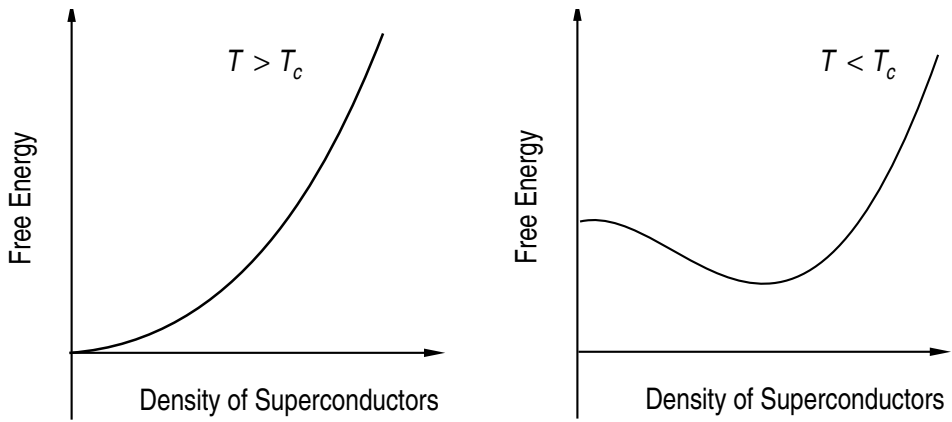  scaled 800}}
	\caption{Ginzburg-Landau description of the superconducting phase 
	transition.}
	\protect\label{fig1}
\end{figure}

Above the critical temperature for the onset of superconductivity, 
the parameter $\alpha$ is positive and the free energy of the substance 
is supposed to be an increasing function of the density of 
superconducting carriers, as shown in Figure \ref{fig1}(a).  The state of minimum 
energy, the vacuum state, then corresponds to a purely resistive flow, 
with no superconducting carriers active.
Below the critical temperature, $\alpha$ is negative and the free energy 
is minimized when $\psi = \psi_0 \ne 0$, as illustrated in Figure \ref{fig1}(b).

This is a nice cartoon description of the superconducting phase 
transition, but there is more.  In an applied magnetic field $\vec{H}$, 
the free energy is
\begin{equation}
G_{\rm super}(\vec{H}) = G_{\rm super}(0) + \frac{\vec{H}^2}{8\pi} + 
\frac{1}{2m^\star}|-i\hbar\nabla\psi-(e^\star/c)\vec{A}\psi|^2 
\;\;,
\end{equation}
where $e^\star$ and $m^\star$ are the charge ($-2$ units) and effective 
mass of the superconducting carriers.  In a weak, slowly varying field 
$\vec{H} \approx 0$, when we can approximate $\psi\approx\psi_0$ and 
$\nabla\psi\approx 0$, the usual variational analysis leads to the 
equation of motion,
\begin{equation}
\nabla^2\vec{A}-\frac{4\pi e^\star}{m^\star c^2}\abs{\psi_0}^2\vec{A} = 
0\;\;,
\end{equation}
the wave equation of a massive photon.  In other words, the photon 
acquires a mass within the superconductor.  This is the origin of the 
Meissner effect, the exclusion of a magnetic field from a 
superconductor.  More to the point, for our purposes, it shows how a 
symmetry-hiding phase transition can lead to a massive gauge boson.

To give masses to the intermediate bosons of the weak interaction, we 
take advantage of a relativistic generalization of the Ginzburg-Landau 
phase transition known as the Higgs mechanism.\cite{5}  We introduce auxiliary 
scalar fields, with gauge-invariant interactions among themselves and 
with the fermions and bosons of the electroweak theory.  We then arrange 
their self-interactions so that the vacuum state corresponds to a 
broken-symmetry solution.  As a result, the $W$ and $Z$ bosons acquire 
masses, as auxiliary scalars assume the role of the third 
(longitudinal) degrees of freedom of what had been massless gauge 
bosons.  The quarks and leptons acquire masses as well, from their Yukawa 
interactions with the scalars.  Finally, there remains as a vestige of 
the spontaneous breaking of the symmetry a massive, spin-zero particle, 
the Higgs boson.  Though what we take to be the work of the Higgs boson 
is all around us, the Higgs particle itself has not yet been observed.

It is remarkable that the resulting theory has been tested at distances 
ranging from about $10^{-17}~\hbox{cm}$ to about $4\times 
10^{20}~\hbox{cm}$, especially when we consider that classical 
electrodynamics has its roots in the tabletop experiments that gave us 
Coulomb's law.  These basic ideas were modified in response to the 
quantum effects observed in atomic experiments.  High-energy physics 
experiments both inspired and tested the unification of weak and 
electromagnetic interactions.  At distances longer than common 
experience, electrodynamics---in the form of the statement that the 
photon is massless---has been tested in measurements of the magnetic 
fields of the planets.  With additional assumptions, the observed 
stability of the Magellanic clouds provides evidence that the photon is 
massless over distances of about $10^{22}~\hbox{cm}$.

\section{A Higgs Boson Must Exist}
\begin{figure}[tb]
	\centerline{\BoxedEPSF{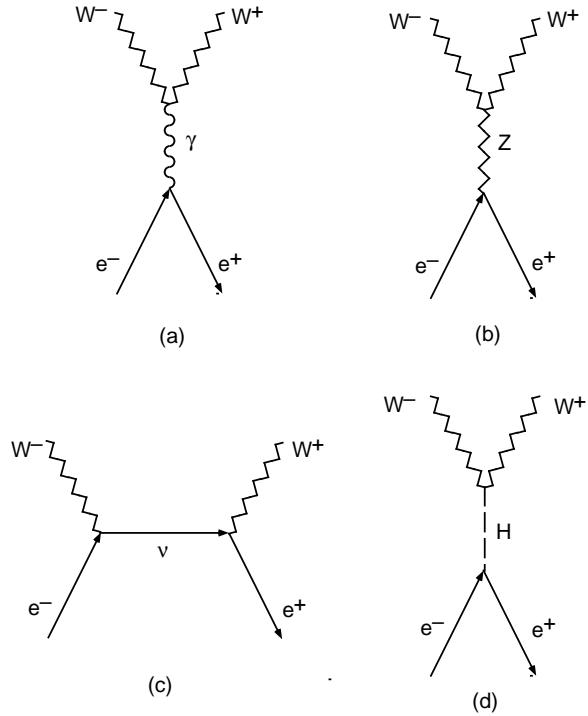  scaled 700}}
	\caption{Lowest-order contributions to the $e^+e^- \rightarrow 
	W^{+}W^{-}$ scattering amplitude.}
	\protect\label{fig:eeWW}
\end{figure}

How can we be sure that a Higgs boson, or something very like it, will be 
found? One 
path to the \emph{theoretical} discovery of the Higgs boson
involves its role in the cancellation of 
high-energy divergences. An illuminating example is provided by the 
reaction
\begin{equation}
	e^+e^- \to W^+W^-,
\end{equation}
which is described in lowest order by the four 
Feynman graphs in Figure \ref{fig:eeWW}. The contributions of the direct-channel 
$\gamma$- and $Z^0$-exchange diagrams 
of Figs.~\ref{fig:eeWW}(a) and (b) cancel the leading divergence in the $J=1$ 
partial-wave amplitude of 
the neutrino-exchange diagram in Figure~\ref{fig:eeWW}(c).
However, the $J=0$ partial-wave amplitude, which exists in this 
case because 
the electrons are massive and may therefore be found in the ``wrong'' 
helicity state, grows as $s^{1/2}$ for the production of longitudinally 
polarized gauge bosons. The resulting divergence is precisely cancelled by 
the Higgs boson graph of Figure~\ref{fig:eeWW}(d). If the Higgs boson did not exist, 
something else would have to play this role. From the point of view 
of $S$-matrix analysis, the Higgs-electron-electron coupling must be 
proportional to the electron mass, because ``wrong-helicity'' amplitudes 
are always proportional to the fermion mass.

Let us underline this result.
If the gauge symmetry were unbroken, there would be 
no Higgs boson, no longitudinal gauge bosons, and no extreme divergence 
difficulties. But there would be no viable low-energy phenomenology
 of the 
weak interactions. The most severe divergences of individual diagrams 
are eliminated by the gauge 
structure of the couplings among gauge bosons and leptons. A lesser, but 
still potentially fatal, divergence arises because the electron has 
acquired mass---because of the Higgs mechanism. Spontaneous symmetry 
breaking provides its own cure by supplying a Higgs boson to remove the 
last divergence. A similar interplay and compensation must exist in any 
satisfactory theory.
\begin{quotation}
	{\small \textbf{Problem: }
Carry out the computation of the amplitudes for the 
reaction $e^+ e^- \rightarrow W^+ W^-$ described above, 
retaining the electron mass.  Verify the role of the Higgs boson in the 
cancellation of divergences.  (\textit{Nota bene:} If you do only one 
gauge-theory calculation in your life, this should be the one!)	
	}
\end{quotation}


\section{Why is the Electroweak Scale Small?}
The $SU(2)_L \otimes U(1)_Y$ electroweak theory does not explain how the 
scale of electroweak symmetry breaking is maintained in the presence 
of quantum corrections.  The problem of the scalar sector can be 
summarized neatly as follows.\cite{10}  The Higgs potential is
\begin{equation}
      V(\phi^\dagger \phi) = \mu^2(\phi^\dagger \phi) +
\abs{\lambda}(\phi^\dagger \phi)^2 \;.
\end{equation}
With $\mu^2$ chosen to be less than zero, the electroweak symmetry is 
spontaneously broken down to the $U(1)$ of electromagnetism, as the 
scalar field acquires a vacuum expectation value that is fixed by the low-energy
phenomenology, 
\begin{equation}
	\vev{\phi} = \sqrt{-\mu^2/2|\lambda|} \equiv (G_F\sqrt 8)^{-1/2}
		\approx 175 {\rm \;GeV}\;.
		\label{hvev}
\end{equation}

Beyond the classical approximation, scalar mass parameters receive 
quantum corrections from loops that contain particles of spins 
$J=1, 1/2$, and $0$:
\begin{equation}
\BoxedEPSF{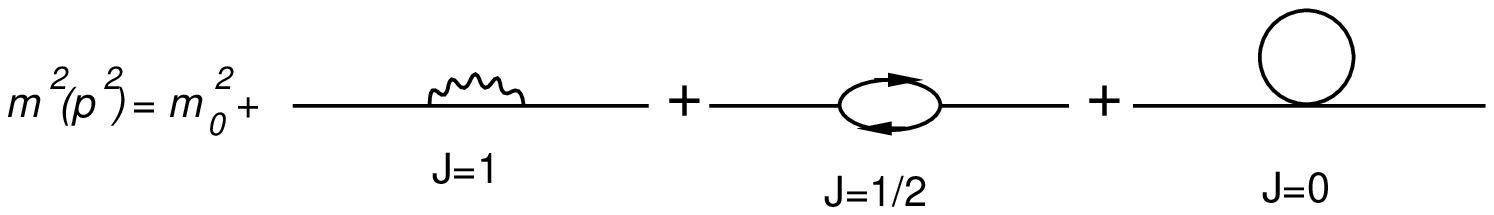  scaled 800}
\label{loup}
\end{equation}
The loop integrals are potentially divergent.  Symbolically, we may 
summarize the content of \eqn{loup} as
\begin{equation}
	m^2(p^2) = m^2(\Lambda^2) + Cg^2\int^{\Lambda^2}_{p^2}dk^2 
	+ \cdots \;,
	\label{longint}
\end{equation}
where $\Lambda$ defines a reference scale at which the value of 
$m^2$ is known, $g$ is the coupling constant of the theory, and the 
coefficient $C$ is calculable in any particular theory.  
Instead of dealing with the relationship between observables and 
parameters of the Lagrangian, we choose to describe the variation of 
an observable with the momentum scale.  In order for the mass shifts 
induced by radiative corrections to remain under control (\ie , not to 
greatly exceed the value measured on the laboratory scale), either 
\begin{itemize}
\item $\Lambda$ must be small, so the range of integration is not 
enormous, or 
\item new physics must intervene to cut off the integral.
\end{itemize}

If the fundamental interactions are described by an 
$SU(3)_c\otimes SU(2)_L\otimes U(1)_Y$ gauge symmetry, \ie, by quantum
chromodynamics and the electroweak theory, then the 
natural reference scale is the Planck mass,
\begin{equation}
	\Lambda \sim M_{\rm Planck} \approx 10^{19} {\rm \; GeV}\;.
\end{equation}
In a unified theory of the strong, weak, and electromagnetic 
interactions, the natural scale is the unification scale,
\begin{equation}
	\Lambda \sim M_U \approx 10^{15}\hbox{-}10^{16} {\rm \; GeV}\;.
\end{equation}
Both estimates are very large compared to the scale of electroweak 
symmetry breaking \eqn{hvev}.  We are therefore assured that new physics must 
intervene at an energy of approximately 1~TeV, in order that the 
shifts in $m^2$ not be much larger than \eqn{hvev}.

Only a few distinct scenarios for controlling the 
contribution of the integral in \eqn{longint} can be envisaged.  The 
supersymmetric solution is especially elegant.  Exploiting the fact 
that fermion loops contribute with an overall minus sign (because of 
Fermi statistics), supersymmetry balances the contributions of fermion 
and boson loops.  In the limit of unbroken supersymmetry, in which the 
masses of bosons are degenerate with those of their fermion 
counterparts, the cancellation is exact:
\begin{equation}
	\sum_{{i={\rm fermions \atop + bosons}}}C_i\int dk^2 = 0\;.
\end{equation}
If the supersymmetry is broken (as it must be in our world), the 
contribution of the integrals may still be acceptably small if the 
fermion-boson mass splittings $\Delta M$ are not too large.  The 
condition that $g^2\Delta M^2$ be ``small enough'' leads to the 
requirement that superpartner masses be less than about 
$1\tevcc$.

A second solution to the problem of the enormous range of integration in 
\eqn{longint} is offered by theories of dynamical symmetry breaking such as 
technicolor. In technicolor models, the Higgs boson is composite, and 
new physics arises on the scale of its binding, $\Lambda_{TC} \simeq 
O(1~{\rm TeV})$. Thus the effective range of integration is cut off, and 
mass shifts are under control.

A third possibility is that the gauge sector becomes strongly 
interacting. This would give rise to $WW$ resonances, multiple 
production of gauge bosons, and other new phenomena at energies of 1 TeV 
or so.  It is likely that a scalar bound state---a quasi-Higgs 
boson---would emerge with a mass less than about $1\tevcc$.

We cannot avoid the conclusion that some new physics must occur on 
the \onetev.

\section{Triviality of Scalar Field Theory}
The electroweak theory itself provides another reason to expect that 
discoveries will not end with the Higgs boson.  Scalar field theories 
make sense on all energy scales only if they are noninteracting, or 
``trivial.''\cite{15}  The vacuum of quantum field theory is a dielectric 
medium that screens charge.  Accordingly, the effective charge is a 
function of the distance or, equivalently, of the energy scale.  This is 
the famous phenomenon of the running coupling constant.

In $\lambda\phi^4$ theory (compare the interaction term in the Higgs 
potential), it is easy to calculate the variation of the coupling 
constant $\lambda$ in perturbation theory by summing bubble graphs like 
this one:
\begin{equation}
\BoxedEPSF{Bulle.epsf  scaled 600}\;\;\;\;.
\end{equation} \vphantom{{\LARGE |}}The coupling constant $\lambda(\mu)$ on a physical scale $\mu$ 
is related 
to the coupling constant on a higher scale $\Lambda$ by
\begin{equation}
\frac{1}{\lambda(\mu)} = \frac{1}{\lambda(\Lambda)} + 
\frac{3}{2\pi^2}\log{\left(\Lambda/\mu\right)}\;\;.
\label{rng}
\end{equation}
This perturbation-theory result is reliable only when $\lambda$ is small, 
but lattice field theory allows us to treat the strong-coupling regime.

In order for the Higgs potential to be stable (\ie, for the energy of the 
vacuum state not to race off to $-\infty$), $\lambda(\Lambda)$ must not 
be negative.  Therefore we can rewrite \eqn{rng} as an inequality,
\begin{equation}
\frac{1}{\lambda(\mu)} \ge 
\frac{3}{2\pi^2}\log{\left(\Lambda/\mu\right)}\;\;. 
\end{equation}
This gives us an {\em upper bound},
\begin{equation}
\lambda(\mu) \le 
2\pi^2/3\log{\left(\Lambda/\mu\right)}\;\;,
\label{upb}
\end{equation}
on the coupling strength at the physical scale $\mu$.
If we require the theory to make sense to arbitrarily high energies---or 
short distances---then we must take the limit $\Lambda\rightarrow\infty$ 
while holding $\mu$ fixed at some reasonable physical scale.  In this 
limit, the bound \eqn{upb} forces $\lambda(\mu)$ to zero.  The scalar field 
theory has become free field theory; in theorist's jargon, it is trivial.

We can rewrite the inequality \eqn{upb} as a bound on the Higgs-boson mass.  
Rearranging and exponentiating both sides gives the condition
\begin{equation}
\Lambda \le \mu \exp{\left(\frac{2\pi^2}{3\lambda(\mu)}\right)}\;\;.
\end{equation}
Choosing the physical scale as $\mu=M_H$, and remembering that, before 
quantum corrections,
\begin{equation}
M_H^2 = 2\lambda(M_H)v^2\;\;,
\end{equation}
where $v=(G_F\sqrt{2})^{-1/2}\approx 246~\hbox{GeV}$ is the vacuum 
expectation value of the Higgs field times $\sqrt{2}$, we find that
\begin{equation}
\Lambda \le M_H\exp{\left(\frac{4\pi^2v^2}{3M_H^2}\right)}\;\;.
\end{equation}
For any given Higgs-boson mass, there is a maximum energy scale 
$\Lambda^\star$ at which the theory ceases to make sense.  The 
description of the Higgs boson as an elementary scalar is at best an 
effective theory, valid over a finite range of energies.

If the Higgs boson is relatively light---which would itself require 
explana\-tion---then the theory can be self-consistent up to 
very high energies.  If the electroweak theory is to make sense all the 
way up to a unification scale $\Lambda^\star = 10^{16}~\hbox{GeV}$, then 
the Higgs boson must weigh less than $170~\hbox{GeV}\!/\!c^2$.\cite{16}

This perturbative analysis breaks down when the Higgs-boson mass 
approaches $1~\hbox{TeV}\!/\!c^2$ and the interactions become strong.  
Lattice analyses~\cite{17} indicate that, for the theory to make sense up 
to a few TeV, the mass of the Higgs boson can be no more than about 
$800~\hbox{GeV}\!/\!c^2$.  Another way of putting this result is that, if 
the elementary Higgs boson takes on the largest mass allowed by 
perturbative unitarity arguments, the electroweak theory will be living 
on the brink of instability.

\section{New Physics Nearby}

\subsection{Supersymmetry}
Since we have profited in Ma\'{o} from Gian Giudice's excellent 
lecture series on supersymmetric models, I will content myself with a 
few general remarks about the motivation for supersymmetry on the 
electroweak scale, and its connection with string theory.\cite{lykken}

One of the best phenomenological motivations for supersymmetry on the 
\onetev\ is that the minimal supersymmetric extension of the standard 
model so closely approximates the standard model itself.  This is a 
nontrivial property of new physics beyond the standard model, and a 
requirement urged on us by the unbroken quantitative success of the 
established theory.  On the aesthetic---or theoretical---side, supersymmetry 
is the maximal---indeed, unique---extension of Poincar\'{e} invariance.  
It also offers a path to the incorporation of gravity, since local 
supersymmetry leads directly to supergravity.  As a practical matter, 
supersymmetry on the \onetev\ offers a solution to the naturalness 
problem, and allows a fundamental scalar to exist at low energies.

When we combine supersymmetry with unification of the fundamental 
forces, we obtain a satisfactory prediction for the weak mixing 
parameter, $\sin^{2}\theta_{W}$, and a simple picture of 
coupling-constant unification.  Adding an assumption of universality, 
we are led naturally to a picture in which the top mass is linked 
with the electroweak scale, so that $m_{t} \approx v/\sqrt{2}$.  
Finally, the assumption of $R$-parity leads to a stable lightest 
supersymmetric particle, which is a natural candidate for the dark 
matter of the Universe.

Supersymmetry doubles the spectrum of fundamental particles.  We know 
that supersymmetry must be significantly broken in Nature, because the 
electron is manifestly not degenerate in mass with its scalar partner, 
the selectron.  It is interesting to contemplate just how different 
the world would have been if the selectron, not the electron, were 
the lightest charged particle and therefore the stable basis of 
everyday matter.\cite{18params}  If atoms were selectronic, there would be 
no Pauli principle to dictate the integrity of molecules.  As 
Dyson~\cite{dyson} and Lieb~\cite{lieb} demonstrated, transforming 
electrons and nucleons from fermions to bosons would cause all 
molecules to shrink into an insatiable undifferentiated blob.  
Luckily, there is no analogue of chiral symmetry to 
guarantee naturally small squark and slepton masses.  So while 
supersymmetry menaces us with an amorphous death, it is 
likely that a full understanding of supersymmetry will enable us to 
explain why we live in a universe ruled by the exclusion principle.

Many theorists take a step beyond supersymmetry to string theory, the 
only known consistent theory of quantum gravity.\cite{quevedo}  String theory 
aspires to unite all the fundamental interactions in one (and only 
one?) theory with few free parameters.  If successful, this program 
might explain the standard-model gauge group, unified extensions to the
\sm\ gauge symmetry, and the fermion content of the standard model.  
String theory makes two generic predictions for physics beyond the 
standard model: additional $U(1)$ subgroups of the unifying group lead 
to new gauge bosons, and additional colored fermions augment the 
spectrum of fundamental constituents.

In spite of what doubters often say, there \emph{is} experimental support 
for string theory from accelerator experiments.  Superstrings predicted 
gravity in 1974,\cite{joel} and LEP accelerator physicists detected tidal 
forces in 1993.\cite{tides}  What more empirical evidence could one 
demand?

\subsection{Technicolor}
Dynamical symmetry breaking provides a different solution to the 
naturalness problem of the electroweak theory: in technicolor, there 
are no elementary scalars.  We hope that solving the dynamics that 
binds elementary fermions into a composite Higgs boson and other $WW$ 
resonances will bring addition predictive power.  It is worth saying 
that technicolor is a far more ambitious program than global 
supersymmetry.  It doesn't merely seek to finesse the hierarchy 
problem, it aims to predict the mass of the Higgs surrogate.  Against 
the aesthetic appeal of supersymmetry we can weigh technicolor's excellent 
pedigree.  As we have seen in \S\ref{cache},
the Higgs mechanism of the standard model 
is the relativistic generalization of the Ginzburg-Landau description 
of the superconducting phase transi\-tion.  Dynamical symmetry breaking 
schemes---technicolor and its relatives---are inspired by the
Bardeen--Cooper--Schrieffer theory of superconductivity, 
and seek to give a similar microscopic description of electroweak 
symmetry breaking.  

In its simplest form, technicolor makes no prediction for fermion 
masses.  Consequently, technicolor serves as a reminder that there are two problems 
of mass: explaining the masses of the gauge bosons, which demands and 
understanding of electroweak symmetry breaking; and accouting for the 
quark and lepton masses, which requires not only an understanding of 
electroweak symmetry breaking but also a theory of the Yukawa 
couplings that set the scale of fermion masses in the standard model.

The dynamical-symmetry-breaking approach realized in technicolor theories
 is modeled upon our understanding of the superconducting phase 
transition.\cite{varsovie,18} The macroscopic order parameter of the Ginzburg-Landau 
phenomenology corresponds to the wave function of superconducting 
charge carriers. As we have seen in \S\ref{landm} and \ref{cache}, it acquires a nonzero vacuum
expectation value in the 
superconducting state. The microscopic Bardeen-Cooper-Schrieffer 
theory\cite{19} identifies the dynamical origin of the order parameter with 
the formation of bound states of elementary fermions, the Cooper pairs of 
electrons. The basic idea of  technicolor is to replace the 
elementary Higgs boson with a fermion-antifermion bound 
state. By analogy with the superconducting phase transition, the dynamics 
of the fundamental technicolor gauge interactions among technifermions 
generate scalar bound states, and these play the role of the Higgs fields.

The elementary fermions---electrons---and 
the gauge interactions---QED---needed to generate the scalar bound states are 
already present in the case of superconductivity. Could a scheme
 of similar economy account
for the transition that hides the electroweak symmetry?
Consider an $SU(3)_c\otimes SU(2)_L\otimes U(1)_Y$ theory of massless up and 
down quarks. Because the strong interaction is strong, and the electroweak 
interaction is feeble, we may treat the $SU(2)_L\otimes U(1)_Y$ 
interaction as a perturbation. For vanishing quark masses, QCD has an exact 
$SU(2)_L\otimes SU(2)_R$ chiral symmetry. At an energy scale 
$\sim\Lambda_{\mathrm{QCD}},$ the strong interactions become strong, fermion 
condensates appear, and the chiral symmetry is spontaneously broken
to the familiar flavor symmetry:
\begin{equation}
	SU(2)_L\otimes SU(2)_R \to SU(2)_V\;\; .
\end{equation}
 Three Goldstone bosons appear, one for 
each broken generator of the original chiral invariance. These were 
identified by Nambu~\cite{20} as three massless pions.

The broken generators are three axial currents whose couplings to pions are 
measured by the pion decay constant $f_\pi$. When we turn on the 
$SU(2)_L\otimes U(1)_Y$ electroweak interaction, the electroweak gauge 
bosons couple to the axial currents and acquire masses of order $\sim 
gf_\pi$. The massless pions thus disappear from the physical spectrum, 
having become the longitudinal components of the weak gauge bosons. 
Unfortunately, the mass acquired by the 
intermediate bosons is far smaller than required for a successful 
low-energy phenomenology; it is only~\cite{21} $M_W\sim 30~\mev\!/\!c^2$.

The minimal technicolor model of Weinberg~\cite{22} and Susskind~\cite{23} 
transcribes the same ideas from QCD to a new setting.  The 
technicolor gauge group is taken to be $SU(N)_{TC}$ (usually $SU(4)_{TC}$), 
so the gauge interactions of the theory are generated by
\begin{equation}
	SU(4)_{TC}\otimes SU(3)_c \otimes SU(2)_L \otimes U(1)_Y\; .
\end{equation}
The technifermions are a chiral doublet of massless color singlets
\begin{equation}
\begin{array}{cc}
	\left( \begin{array}{c} U \\ D \end{array} \right)_L & U_R, \;
D_R \; .
\end{array}
\end{equation}
With the electric charge assignments $Q(U)=\frac{1}{2}$ and
$Q(D)=-\frac{1}{2}$, the  
theory is free of electroweak anomalies. The ordinary fermions are all 
technicolor singlets. 

In analogy with our discussion of chiral symmetry breaking in QCD, we 
assume that the chiral $TC$ symmetry is broken,
\begin{equation}
	SU(2)_L\otimes SU(2)_R\otimes U(1)_V\to SU(2)_V\otimes U(1)_V\; .
\end{equation}
Three would-be Goldstone bosons emerge. These are the technipions
\begin{equation}
\begin{array}{ccc}
\pi_T^+, & \pi_T^0, & \pi_T^-,
\end{array}
\end{equation}
for which we are free to {\em choose} the technipion decay constant as
\begin{equation}
	F_\pi = \left(G_F\sqrt{2}\right)^{-1/2} = 246\gev\; . \label{FPI}
\end{equation}
This amounts to choosing the scale on which technicolor becomes strong.
When the electroweak interactions are turned on, the technipions become the 
longitudinal components of the intermediate bosons, which acquire masses
\begin{equation}
\renewcommand\arraystretch{1.5}
\begin{array}{ccccc}
	M_W^2 & = & g^2F_\pi^2/4 & = & 
{\displaystyle \frac{\pi\alpha}{G_F\sqrt{2}\sin^2\theta_W}} \\
	M_Z^2 & = & \left(g^2+g^{\prime 2}\right)F_\pi^2/4 & = & 
M_W^2/\cos^2\theta_W
\end{array}
\renewcommand\arraystretch{1}
\end{equation}
that have the canonical Standard Model values, thanks to our choice 
(\ref{FPI}) 
of the technipion decay constant.

Technicolor shows how the generation of intermediate boson masses could 
arise without fundamental scalars or unnatural adjustments of parameters. 
It thus provides an elegant solution to the naturalness problem of the 
Standard Model. However, it has a major deficiency: it offers no 
explanation for the origin of quark and lepton masses, because no Yukawa 
couplings are generated between Higgs fields and quarks or leptons. 

A possible approach to the problem of quark and lepton masses is suggested 
by ``extended technicolor'' models and their modern extensions, ``walking 
technicolor'' models.\cite{24}  Technicolor implies a number of spinless technipions 
with masses below the technicolor scale of about 1~\tev. Some of these, the 
color singlet, technicolor singlet particles, should be relatively light.
The colored technipions and technivector mesons may just be  
accessible to experiments at the Tevatron, but a 
thorough investigation awaits experiments on the \onetev.\cite{elworm}

\subsection{Implications of Heavy Top}
The great mass of the top quark is suggestive for both the 
supersymmetry and dynamical symmetry breaking approaches.  Within the 
framework of supersymmetry, the heavy top quark encourages the belief 
in low-scale supersymmetry, and suggests that the discovery of 
supersymmetry may be at hand, either at LEP2 or at the Tevatron 
Collider.  To me, the proximity of the top mass to the scale of 
electroweak symmetry breaking argues that the two problems of 
mass---mass for the gauge bosons and mass for the fermions---may be 
one.  In other words, the heavy top makes it less likely that the 
question of flavor can be postponed, and more likely that it is of a 
piece with the problem of electroweak symmetry breaking.  This linkage 
is (almost) surely true in any dynamical symmetry breaking scheme, and 
I find it an appealing conclusion in general.  Because of the heavy 
top, I am now optimistic that exploring the \onetev\ will illuminate 
the flavor question as well as electroweak symmetry breaking.

\subsection{New Strong Dynamics}
If new strong dynamics at the \onetev\ replaces the elementary Higgs 
boson of the standard model, then it is reasonable to expect 
low-energy signals of the new dynamics.  There are two interesting 
options to explore:
\begin{itemize}
	\item  Replace the Higgs sector only with a composite.  This is the 
	approach followed in Technicolor and its generalizations to 
	``walking'' technicolor or two-scale technicolor, topcolor, and 
	topcolor-assisted technicolor.

	\item  The quarks and leptons are composite, as well as the Higgs 
	boson, on a scale $\Lambda^{\star}\approx$ a few TeV.  The dynamics 
	of a composite model of quarks and leptons is quite unlike that of 
	QCD, for the massless particle of the theory must be the fermions, 
	not the analogues of pions.  If quarks and leptons are composite, we 
	might hope to gain an understanding of generations and of fermion 
	masses.
\end{itemize}
In the following Section, I briefly review what is known---and 
desired---
about composite models of our ``elementary'' particles, the quarks and 
leptons.

\section{Composite Quarks and Leptons?}
Throughout these lectures, we have assumed the quarks and leptons to be 
elementary point particles. This is consistent with the experimental 
observations to date that the ``size'' of quarks and leptons is bounded 
from above by
\beq
	R<10^{-17}~{\rm cm} \; .
	\label{compr}
\eeq
Indeed, the identification of quarks and leptons as elementary particles 
(whether that distinction holds at all distance scales or only the regime 
we are now able to explore) is an important ingredient in the simplicity of 
the standard model.

We may nevertheless wish to entertain the possibility that the quarks and 
leptons are themselves composites of some still more fundamental 
structureless particles, for the followeing reasons:
\begin{itemize}
\item The proliferation of ``fundamental'' fermions
\beq
\begin{array}{cccc}
\left( \begin{array}{c} u \\ d \end{array} \right)_L &
	\left(\begin{array}{c} c \\ s \end{array} \right)_L &
	\left(\begin{array}{c} t \\ b \end{array} \right)_L &
	u_R, d_R, s_R, c_R, b_R, t_R \\
 & & & \\
\left( \begin{array}{c} \nu_e \\ e \end{array} \right)_L &
	\left(\begin{array}{c} \nu_\mu \\ \mu \end{array} \right)_L &
	\left(\begin{array}{c} \nu_\tau \\ \tau \end{array} \right)_L &
	e_R, \mu_R, \tau_R 
		\end{array} 
\eeq
and the repetition of generations.
\item The complex pattern of masses and angles suggests they may not be 
fundamental parameters. 
\item Hints of a new strong interaction (Technicolor) and the resulting 
composite scalar particles.
\end{itemize}
To this we may add the most potent question of all, Why not?
\subsection{A Prototype Theory of Composite Quarks or Leptons}
Building on our knowledge of gauge theories for the interactions of 
fundamental fermions, we imagine~\cite{68} a set of massless, pointlike, spin-1/2 
{\em preons} carrying the charge of a new gauge interaction called {\em 
metacolor}. The metacolor interaction arises from a gauge symmetry 
generated by the group ${\cal G}$. We 
assume that the metacolor interaction is asymptotically free and infrared 
confining. Below the characteristic energy scale $\Lambda^*$, the metacolor 
interaction become strong (in the sense that $\alpha_M(\Lambda^{*2})\approx 1$)
and binds the preons into metacolor-singlet states including the observed 
quarks and leptons. In this way, the idea of composite quarks and leptons 
may be seen as a natural extension of the technicolor strategy for 
composite Higgs scalars.

We expect from the small size of the quarks and leptons that the 
characteristic energy scale for preon confinement must be quite large,
\beq
	\Lambda^* \gtap 1/R \gtap 1\tev \;\;.
	\label{complow}
\eeq
On this scale, the quarks and leptons are effectively massless. This is the 
essential fact that a composite theory of quarks and leptons must explain: 
the quarks and leptons are both small and light.

In general, it is the scale $\Lambda^*$ that determines the masses of 
composite states. However, there are special circumstances in which some 
composite states will be exactly or approximately massless compared to the 
scale $\Lambda^*$. The Goldstone theorem~\cite{69} asserts that a massless spin-zero particle 
arises as a consequence of the spontaneous breakdown of a continuous global 
symmetry. We have already seen examples of this behavior in the 
small masses of the color-singlet technipions, which arise as Goldstone 
bosons when the chiral symmetry of tecchnicolor is spontaneously broken. 

't~Hooft noted that under certain special conditions, confining theories 
that possess global chiral symmetries may lead to the existence of 
massless composite fermions when the chiral symmetries are not 
spontaneously broken. The key to this observation is the anomaly 
condition~\cite{70} which constrains the pattern of chiral symmetry breaking 
and the spectrum of light composite fermions:
\begin{quotation}
For any conserved global (flavor) current, the same anomaly must arise from 
the fundamental preon fields and from the ``massless'' physical states.
\end{quotation}
The existence of an anomaly therefore implies a massless physical state 
associated with the anomalous charge $Q$. If the (global) chiral or flavor symmetry 
respected by the preons is broken down when the metacolor interaction 
becomes strong as
\beq
	G_f\to S_f\subseteq G_f~{\rm at}~\Lambda^*\;\;,
\eeq
then the consistency condition can be satisfied in one of two ways:
\begin{itemize}
\item If the anomalous charge $Q \not\in S_f$, so the global symmetry which 
has the anomaly is spontaneously broken, then a Goldstone boson arises with 
specified couplings to the anomaly;
\item If instead $Q\in S_f$, so that the anomalous symmetry remains 
unbroken when metacolor becomes strong, then there must be massless, 
spin-$\cfrac{1}{2}$ fermions in the physical spectrum which couple to $Q$ and 
reproduce the anomaly as given by the preons.
\end{itemize}

The anomaly conditions thus show how massless fermions might arise as 
composite states in a strongly interacting gauge theory.
In analogy with the case of the pions, we may then 
suppose that a small bare mass for the preons, or preon electroweak 
interactions that explicitly break the chiral symmetries, can account for 
the observed masses of quarks and leptons. However, there is as yet no realistic 
model of the quark and lepton spectrum. It is natural to ask whether the 
repeated pattern of generations might be an excitation spectrum. The answer 
seems clearly to be No. For a strong gauge interaction, all the excitations 
should occur at a scale $\Lambda^*$ and above.

The scenario that emerges from this rather sketchy discussion of composite 
models is that all quarks and leptons are massless in some approximation. 
Generations arise not from excitations, but because of symmetries coupled 
with the anomaly condition. All masses and mixings arise because of 
symmetry breaking not associated with the composite strong force. This is a 
promising outcome on two out of three counts: We may hope for some insight 
into the near masslessness of quarks and leptons, and into the meaning of 
generations, but the origin of mass and mixings seems as mysterious as ever.
\subsection{Manifestations of Compositeness}
The classic test for substructure is to search for form factor effects, or 
deviations from the expected pointlike behavior in gauge-boson propagators 
and fermion vertices.\cite{71} Such deviations would occur in any composite 
model, at values of $\sqrt{\shat}\gg\Lambda^*$, for example as a 
consequence of vector meson dominance. In a favored parametrization of this 
effect, the gauge field propagator is modified by a factor
\beq
	F(Q^2) = 1+Q^2/\Lambda^{*2}\;\;,
\eeq
where $Q$ is the four-momentum carried by the gauge field. Measurements of 
the reactions
\beq
	e^+e^- \to \left\{ \begin{array}{c}  q\qbar \\
		\ell^+\ell^- \end{array} \right.
		\label{compreac}
\eeq
on and off the $Z^{0}$
yield limits on the compositeness scale which translate into the bound on 
fermion size given in \eqn{compr}.

Many other tests of compositeness can be carried out in the study at low    
energies of small effects or rare transitions sensitive to virtual 
processes. For example, if a composite fermion $f$ is naturally light 
because of 't~Hooft's mechanism, there will arise a contribution to its 
anomalous magnetic moment of order~\cite{72} $(m_f/\Lambda^*)^2$. The close 
agreement~\cite{73} between the QED prediction and the measured value of
$(g-2)_\mu$ implies that
\beq
\Lambda^* \gtap 670\gev
\eeq
for the muon. This is the only constraint on $\Lambda^*$ from anomalous 
moments that improves on the limits from the reactions \eqn{compreac}. Within 
specific models, very impressive bounds on the compositeness scale may be 
derived from the absence of flavor-changing neutral current transitions.

The observations at PEP, PETRA, and TRISTAN lead to
\begin{equation}
	\Lambda^{*} > 1.7\hbox{ to }4.5\tev
\end{equation}
for individual channels, and to
\begin{equation}
	\Lambda^{*} > 2.9\hbox{ to }5.4\tev
\end{equation}
when $e,\mu,\tau$ are combined.\cite{kroha}  LEP measurements also 
lead to bounds of several TeV.\cite{pdgcomp}  At LEP~II, $1\hbox{ 
fb}^{-1}$ of running should yield sensitivity up to about $7\tev$.

At energies below those for which form factor effects become 
characteristic, \ie, for
\beq 
	\sqrt{\shat} \sim~{\rm few~times}~\Lambda^*,
\eeq
we may anticipate resonance formation and multiple production. The latter 
might well include reactions such as 
\beq
	u\overline{u}\to \left\{ \begin{array}{c}
		u\overline{u}u\overline{u} \\
		u\overline{u}e\overline{e} \\
		q^*\qbar^* \end{array} \right. \;\; ,
\eeq
\etc\ In some ways, these would be the most direct and dramatic 
manifestations of compositeness.

At energies small compared to the compositeness scale, the interaction 
between bound states is governed by the finite size of the bound states, 
by the radius $R$. Because the interactions are strong only within this 
confinement radius, the cross section for scattering composite particles 
at low energies should be essentially geometric,
\beq
	\sigma\sim 4\pi R^2 \sim 4\pi/\Lambda^{*2} \;.
\eeq
Regarded instead in terms of the underlying field theory, the low energy 
interaction will be an effective four-fermion interaction, mediated by the 
exchange of massive bound states of preons. When
\beq
	\sqrt{\shat} \ll \Lambda^*,
	\label{complim}
\eeq
the resulting interaction will be a contact term, similar to the low-energy 
limit of the electroweak theory. The general form of the contact 
interaction will be 
\beq
	{\cal L}_{\mathrm{contact}}\sim \frac{g_{\mathrm{Metacolor}}^2}{M_V^2}\cdot
\overline{f}_4 \gamma_\mu f_2 \overline{f}_3 \gamma^\mu f_1 \; .
\label{contact}
\eeq
Identifying $M_V\approx \Lambda^*$ and
\beq
	g_M^2/4\pi =1 \; ,
\eeq
we see that this interaction reproduces the expected geometrical size of 
the cross section in the limit \eqn{complim}.
\subsection{Signals for Compositeness in $p^\pm p$ Collisions}
The flavor-diagonal contact interactions symbolized by \eqn{contact} will modify 
the cross sections for $ff$ elastic scattering. If in the standard model 
this process is controlled by a gauge coupling $\alpha_f \ll 1$, then the 
helicity-preserving pieces of the contact interaction give rise to 
interference terms in the integrated cross section for $ff$ scattering that 
are of order~\cite{74}
\beq
	\frac{\shat}{\Lambda^{*2}}\cdot\frac{g^2}{4\pi\alpha_f} \equiv 
\frac{\shat}{\alpha_f\Lambda^{*2}}
\eeq
relative to the standard model contribution. This modification to the 
conventional expectation is far more dramatic than the anticipated 
$O(\shat/\Lambda^{*2})$ form factor effects. The direct contact term itself 
will dominate for (sub)energies satisfying
\beq
	\shat\gtap\alpha_f\Lambda^{*2}\;\; .
\eeq
The approximation that the composite interactions can be represented by 
contact terms can of course only be reasonable when \eqn{complim} is satisfied.

Although various flavor-changing contact interactions can be tuned away in 
particular models (and must be, in many cases, to survive experimental 
constraints), the flavor-diagonal contact interactions that originate in 
the exchange of common preons must in general survive. This suggests a 
strategy for testing the idea of compositeness:
\begin{quotation}
Consider only four-fermion interactions that are flavor-preserving and 
respect the $SU(3)_c\otimes SU(2)_L\otimes U(1)_Y$ gauge symmetry of the 
standard model.
\end{quotation}
These are unavoidable in a theory capable of producing massless fermionic 
bound states. Three cases are to be considered:
\begin{itemize}
\item electron compositeness;
\item quark compositeness;
\item common lepton-quark compositeness.
\end{itemize}
The second and third, which can be attacked effectively in hadron-hadron 
collisions, will be our concern here.

In the case of quark-quark scattering, we look for deviations from the 
consequences of QCD for the production of hadron jets. The most general contact interactions that respect the gauge symmetry of 
the standard model, involve only up and down quarks, and are helicity 
preserving, involve ten independent terms. Let us consider
 one of these as an example of the phenomena to be 
anticipated in a composite world:
\beq
	{\cal L}_{\mathrm{contact}}^{(0)} = \eta_0\cdot 
\frac{g^2}{2\Lambda^{*2}} \qbar_L\gamma^\mu q_L\qbar_L\gamma_\mu q_L\;,
\label{contact2}
\eeq
where $g^2/4\pi\equiv 1$ and $\eta_0=\pm 1$. This interaction modifies the 
amplitudes for the transitions
\def\ubar{\overline{u}}
\def\dbar{\overline{d}}
\beq
\begin{array}{c}
\begin{array}{cc} u\ubar\to u\ubar & d\dbar\to d\dbar \end{array} \\
\begin{array}{cccc} uu\to uu & dd\to dd & \ubar\ubar\to\ubar\ubar & 
\dbar\dbar\to\dbar\dbar \end{array} \\
	u\ubar\to d\dbar \\
\begin{array}{ccc} ud\to ud & u\dbar\to u\dbar & \ubar d\to \ubar d 
\end{array} \\
	\ubar\dbar\to\ubar\dbar   \end{array}
\eeq
but has no effect on processes involving gluons.

It is convenient to write the differential cross section for the 
parton-parton scatteing process as 
\beq
	\frac{d\hat{\sigma}(ij\to i^\prime j^\prime)}{d\that} = 
\frac{\pi}{\shat^2}\left| A(ij\to i^\prime j^\prime)\right|^2\;.
\eeq
Then in the presence of a contact term \eqn{contact2} the squares of amplitudes are
\begin{eqnarray}
\left| A(ud\to ud)\right|^2 & = & \left| A(u\dbar\to u\dbar) 
\right|^2 \nonumber\\
 & = & \left| A(\ubar d\to \ubar d) \right |^2 = \left| 
A(\ubar\dbar\to\ubar\dbar)\right|^2 \\
 & = & \frac{4}{9}\alpha_s^2(Q^2)\frac{\shat^2+\uhat^2}{\that^2} + 
\left[\frac{\eta_0\uhat}{\Lambda^{*2}}\right]^2 \nonumber\;\;;
\end{eqnarray}
\begin{eqnarray}
\left| A(u\ubar\to d\dbar)\right|^2 & = & \left| A(d\dbar\to 
u\ubar)\right|^2 \nonumber \\
 & = & \frac{4}{9}\alpha_s^2(Q^2)\frac{\that^2+\uhat^2}{\uhat^2} + 
\left[\frac{\eta_0\uhat}{\Lambda^{*2}}\right]^2 \;\;;
\end{eqnarray}
\begin{eqnarray}
\left| A(u\ubar\to u\ubar)\right|^2 & = & \left| A(d\dbar\to 
d\dbar)\right|^2 \nonumber \\
 & = & \frac{4}{9}\alpha_s^2(Q^2)\left[\frac{\uhat^2+\shat^2}{\that^2} + 
\frac{\uhat^2+\that^2}{\shat^2} - \frac{2\uhat^2}{3\shat\that}\right] \\
 & & +\frac{8}{9}\alpha_s(Q^2)\frac{\eta_0}{\Lambda^{*2}}\left[ 
\frac{\uhat^2}{\that} + \frac{\uhat^2}{\shat} \right] + \frac{8}{3} 
\left[\frac{\eta_0\uhat}{\Lambda^{*2}}\right]^2\nonumber\;\;;
\end{eqnarray}
\begin{eqnarray}
\left| A(uu\to uu)\right|^2 & = & \left| A(dd\to dd)\right|^2\nonumber\\
 & = & \left| A(\ubar\ubar\to\ubar\ubar)\right|^2 = \left| 
A(\dbar\dbar\to\dbar\dbar)\right|^2 \\
 & = & \frac{4}{9}\alpha_s^2(Q^2)\left[\frac{\uhat^2+\shat^2}{\that^2} + 
\frac{\shat^2+\that^2}{\uhat^2} - \frac{2\shat^2}{3\uhat\that}\right] \\
 & & +\frac{8}{9}\alpha_s(Q^2)\frac{\eta_0}{\Lambda^{*2}}\left[ 
\frac{\shat^2}{\that} + \frac{\shat^2}{\uhat} \right] + 
\left[\frac{\eta_0}{\Lambda^{*2}}\right]^2\left(\uhat^2+\that^2+\frac{2}{3} 
\shat^2\right)\nonumber\;\;.
\end{eqnarray}
Relative to the QCD terms, the influence of the contact term grows linearly 
with the square $\shat$ of the parton-parton subenergy.  Because the 
contact term modifies the cross section for (anti)quark--(anti)quark 
scattering, its effects are most apparent at the large values of 
$p_{\perp}$ 
for which valence quark interactions dominate the jet cross section.
The search for such enhancements has become a routine part of the 
comparison between jet cross sections and QCD.\cite{CDFcomp}

If quarks and leptons have a common preon constituent, the familiar 
Drell-Yan contribution to dilepton production will be modified by a contact 
term. Whereas the conventional Drell-Yan contribution falls rapidly 
with ${\cal M}$ (because both parton luminosities and the elementary cross 
section do), the cross sections including the contact interaction  
nearly flatten out. The weak dependence upon the effective mass of the 
lepton pair results from the convolution of the rising elementary cross 
section with the falling parton luminosities. There are no conventional 
backgrounds to this signal for quark and lepton substructure.

The contributions of contact terms to dilepton production and jet 
production are comparable. However, in jet production there are large 
incoherent QCD contributions from quark-gluon and gluon-gluon interactions. 
In addition, the standard model cross section for $q\qbar\to\ell^+\ell^-$ is 
smaller than the quark-quark scattering cross section by a factor of order 
$(\alpha_{EM}/\alpha_s)^2$. This accounts for the greater prominence of the 
compositeness signal in dilepton production. 

\begin{quotation}
	{\small \textbf{Problem: }
How would nearby new strong dynamics affect precision electroweak 
measurements?  To be specific, consider the effect of a contact 
interaction, as depicted here\\
	\centerline{\BoxedEPSF{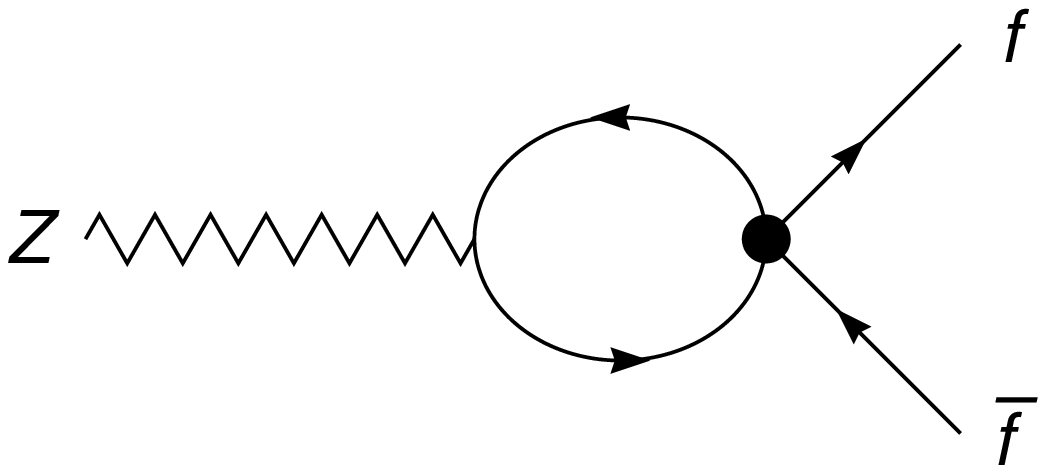  scaled 400}}
on the decay rate for $Z^{0}\rightarrow f\bar{f}$.  Can the excellent 
agreement between experiment and the standard-model predictions for 
$\Gamma(Z^{0}\rightarrow f\bar{f})$ be used to rule out a low 
compositeness scale?  What deviations from the standard model would 
you expect if the compositeness scale were no more than a few TeV?
}
\end{quotation}

\section{The Open Questions}
I close this brief survey of the physics accessible to hadron 
colliders with a catalogue of some of the important questions we 
face.  The central challenge in particle physics is to explore the 
\onetev, and there to elucidate the nature of electroweak symmetry 
breaking.  Because the agent of electroweak symmetry breaking in the 
minimal electroweak theory is the Higgs boson, we may abbreviate this 
problem as the problem of the Higgs sector.  We have recognized the 
significance of the \onetev---the realm of 
electroweak symmetry breaking---for nearly two decades.
Through the development of superconducting magnets, and thanks to
the experience gained in operating high-energy $\bar{p}p$ colliders
at CERN and Fermilab and the evolution of detector architecture from
Mark I at SPEAR up through UA1 and UA2 at CERN and CDF and D\O\
at Fermilab, we now have in hand the technical means to begin our
assault on this frontier of our understanding.  We have also made 
great strides in Monte Carlo tools to design and 
extract physics results from detectors, and in our ability to calculate 
multiparton amplitudes.  But the 
biggest changes in the way we think about the opportunities of 
supercollider physics stem from
\begin{itemize}
	\item  the development of silicon microvertex detectors to tag and 
	measure heavy flavors, and

	\item  the great mass of the top quark.
\end{itemize}

What we seek is an 
understanding of the distinction between the weak and electromagnetic 
interactions in the everyday world, an understanding of the origin of 
masses of the gauge bosons.  We can be confident that 
important clues will be found at the LHC.  

It is less certain that the key to the pattern of quarks and lepton 
masses will be found at a nearby mass scale, but I am encouraged by 
the great mass of the top quark to believe that we may find a common 
resolution of the problems of boson and fermion masses.  At least I 
think it is likely that an intensive study of the top quark will 
provide crucial hints into the puzzle of fermion masses.

Looking beyond the questions we are able to frame with precision, we 
need to understand the origin and meaning of \textsl{CP}-violation, 
the (possibly related) significance of quark-lepton generations, the 
origin of the observed gauge symmetries and, specifically, the reason 
for parity violation in the weak interactions.

By opening up new domains of high energy and short distance, hadron 
colliders lead us to the realm of pure exploration: to the search for 
new forces and new forms of matter, and to test of the notion that 
quarks and leptons might be composite, rather than fundamental.

Let the exploration begin!

\section*{Acknowledgments}
I am grateful to R. Michael Barnett and the Particle Data Group for 
perimission to reproduce Figure \ref{fig:f2}.  I thank Mary Hall Reno 
for prparing Figures \ref{fig:xu} and \ref{fig:momfrac}.

Fermilab is operated by Universities Research 
Association, Inc., under contract DE-AC02-76CHO3000 with the U.S. 
Department of Energy.  I am grateful to Princeton University for warm 
hospitality during the writing of these 
notes.  I thank the organizers and students for a stimulating week in 
Ma\'{o}.


\begin{thebibliography}{99}
\bibitem{toplite} For $m_t \ltap M_W - m_b$, standard top decay 
	($t\rightarrow bW^\star$) is a weak process, which would have been 
	swamped by a (hypothetical) semiweak $t\rightarrow bP^+$ process.  The 
	existence of such a competing decay would have invalidated lower bounds 
	on $m_t$ derived from fruitless searches in the dilepton channel.
\bibitem{dcexp}
	For measurements of the multiplicity in tagged-$b$ 
	events on the $Z^0$ resonance, see R. Akers, \etal\ (OPAL 
	Collaboration), \zp{61}{209}{94}; K. Abe, \etal\ (SLD Collaboration), 
	\prl{72}{3145}{94}.  Tracks arising from $b$-decay are subtracted.
\bibitem{su5} All the important features emerge in an $SU(5)$ unified 
theory that contains the standard-model gauge group $SU(3)_{c}\otimes 
SU(2)_{L} \otimes U(1)_{Y}$.  The final result is more general.
\end{thebibliography}
\end{document}